\documentclass{article}
\usepackage{amsmath,amsbsy,amsfonts}

\begin{document}
\newcommand{\setl[1]}{\setlength{\unitlength}{#1cm}}
\newcommand{\nn}{\nonumber \\}
\newcommand{\eqqref}[1]{\mbox{Eq. (\ref{#1})}}
\newcommand{\eqpref}[1]{\mbox{(Eq. \ref{#1})}}
\newcommand{\eps}{\ensuremath{\varepsilon}}
\newcommand{\me}{\mathrm{e}}
\newcommand{\im}{\ensuremath{\mathrm{i}}}
\newcommand{\bra}[1]{\langle #1 |}
\newcommand{\ket}[1]{| #1 \rangle}
\newcommand{\proj}[1]{\ket{#1}\bra{#1}}
\newcommand{\Bra}[1]{\left\langle #1 \left|}
\newcommand{\Ket}[1]{\right| #1 \right\rangle}
\newcommand{\bigbra}[1] {\big\langle #1\big|}
\newcommand{\bigket}[1] {\big|#1\big\rangle}
\newcommand{\Bigbra}[1] {\Big\langle #1\Big|}
\newcommand{\Biggbra}[1] {\Bigg\langle #1\Bigg|}
\newcommand{\Bigket}[1] {\Big|#1\Big\rangle}
\newcommand{\Biggket}[1] {\Bigg|#1\Bigg\rangle}
\newcommand{\dif}{\ensuremath{\mathrm{d}}}
\newcommand{\difbx}{\dif^3\bx}
\newcommand{\difbk}{\dif^3\bk}
\newcommand{\difbp}{\dif^3\bp}
\newcommand{\intDim}[1]{\int\frac{\dif^D #1}{(2\pi)^D}}
\newcommand{\intD}[1]{\int\frac{\dif^3 #1}{(2\pi)^3}}
\newcommand{\intDD}[1]{\int\frac{\dif^4 #1}{(2\pi)^4}}
\newcommand{\intdinf}[1]{\int_{-\infty}^\infty\frac{\dif #1}{2\pi}}
\newcommand{\intbx}{\int\dif^3\bx}
\newcommand{\bx}{\boldsymbol{x}}
\newcommand{\bxdot}{\dot{\bx}}
\newcommand{\br}{\boldsymbol{r}}
\newcommand{\bk}{\boldsymbol{k}}
\newcommand{\bp}{\bs{\rm{p}}}
\newcommand{\bpi}{\boldsymbol{\pi}}
\newcommand{\bq}{\bs{\rm{q}}}
\newcommand{\bl}{\bs{\rm{l}}}
\newcommand{\bj}{\bs{\rm{j}}}
\newcommand{\bsp}{\bs{\rm{s}}}
\newcommand{\bL}{\bs{\rm{L}}}
\newcommand{\bJ}{\bs{\rm{J}}}
\newcommand{\bI}{\bs{\rm{I}}}
\newcommand{\bF}{\bs{\rm{F}}}
\newcommand{\bS}{\bs{\rm{S}}}
\newcommand{\hatl}{\hat{l}}
\newcommand{\hats}{\hat{s}}
\newcommand{\hatj}{\hat{j}}
\newcommand{\hatbj}{\hat{\bs{\rm{j}}}}
\newcommand{\hatbl}{\hat{\bs{\rm{l}}}}
\newcommand{\hatbs}{\hat{\bs{\rm{s}}}}
\newcommand{\hatL}{\hat{L}}
\newcommand{\hatS}{\hat{S}}
\newcommand{\hatJ}{\hat{J}}
\newcommand{\hatbJ}{\hat{\bs{\rm{J}}}}
\newcommand{\hatbL}{\hat{\bs{\rm{L}}}}
\newcommand{\hatbS}{\hat{\bs{\rm{S}}}}
\newcommand{\hatH}{\hat{H}}
\newcommand{\hrho}{\hat{\rho}}
\newcommand{\balpha}{\ensuremath{\boldsymbol{\alpha}}}
\newcommand{\bgamma}{\bs{\rm{\gamma}}}
\newcommand{\bgd}{\bs{\rm{\gamma}}\bsdot\!}
\newcommand{\bgp}{\bs{\rm{\gamma}}\bsdot\bp}
\newcommand{\bgq}{\bs{\rm{\gamma}}\bsdot\bq}
\newcommand{\bgk}{\bs{\rm{\gamma}}\bsdot\bk}
\newcommand{\gnp}{\gamma^0p_0}
\newcommand{\gnk}{\gamma^0k_0}
\newcommand{\giki}{\gamma^ik_i}
\newcommand{\gipi}{\gamma^ip_i}
\newcommand{\gmu}{\gamma^\mu}
\newcommand{\gnu}{\gamma^\nu}
\newcommand{\gs}{\gamma^\sigma}
\newcommand{\ga}{\gamma}
\newcommand{\wts}{\wt{\gamma}^\sigma}
\newcommand{\wtg}{\wt{\gamma}}
\newcommand{\gt}{\gamma^\tau}
\newcommand{\gb}{\gamma^\beta}
\newcommand{\gn}{\gamma^0}
\newcommand{\gi}{\gamma^i}
\newcommand{\gii}{\gamma_i}
\newcommand{\gj}{\gamma^j}
\newcommand{\gjj}{\gamma_j}
\newcommand{\gE}{\gamma_\rm{E}}
\newcommand{\dd}[1]{\frac{\partial}{\partial #1}}
\newcommand{\Partder}[1]{\frac{\partial }{\partial #1}}
\newcommand{\Partdern}[2]{\frac{\partial^#1 }{\partial#2 ^#1}}
\newcommand{\partder}[2]{\frac{\partial #1}{\partial #2}}
\newcommand{\partdern}[3]{{\frac{\partial^#1 #2}{\partial #3^#1}}}
\newcommand{\Cov}{\mathrm{Cov}}
\newcommand{\eff}{_{\mathrm{eff}}}
\newcommand{\SEp}{_{\mathrm{SEp}}}
\newcommand{\Irred}{\mathrm{Irred}}
\newcommand{\Sep}{\mathrm{Sep}}
\newcommand{\Nonsep}{\mathrm{Nonsep}}
\newcommand{\Irr}{_{\mathrm{Irr}}}
\newcommand{\RL}{_{\mathrm{L}}}
\newcommand{\QL}{_{\mathrm{QL}}}
\newcommand{\sgn}{\mathrm{sgn}}
\newcommand{\etc}{\mathrm{etc}}
\newcommand{\out}{\mathrm{out}}
\newcommand{\iin}{\mathrm{in}}
\newcommand{\inter}{\mathrm{int}}
\renewcommand{\rm}{\mathrm}
\newcommand{\mI}{\mathrm{I}}
\newcommand{\mS}{\mathrm{S}}
\newcommand{\mD}{\mathrm{D}}
\newcommand{\mG}{\mathrm{G}}
\newcommand{\mH}{\mathrm{H}}
\newcommand{\ext}{\mathrm{ext}}
\newcommand{\conn}{\mathrm{conn}}
\newcommand{\Counter}{\mathrm{Counter}}
\newcommand{\Linked}{\mathrm{linked}}
\newcommand{\linked}{\mathrm{linked}}
\newcommand{\mB}{\mathrm{B}}
\newcommand{\mC}{\mathrm{C}}
\newcommand{\SE}{\mathrm{SE}}
\newcommand{\G}{\mathrm{G}}
\newcommand{\sr}{\mathrm{sr}}
\newcommand{\op}{\mathrm{op}}
\newcommand{\opL}{\mathrm{opL}}
\newcommand{\cl}{\mathrm{cl}}
\newcommand{\clC}{\mathrm{clC}}
\newcommand{\bs}{\boldsymbol}
\newcommand{\bdot}{\boldsymbol\cdot}
\newcommand{\bnabla}{\bs{\nabla}}
\newcommand{\Q}{\mathcal{\boldsymbol{Q}}}
\newcommand{\bOm}{\mathcal{\boldsymbol{\Om}}}
\renewcommand{\H}{H}
\newcommand{\A}{A}
\newcommand{\bH}{\bs{\H}}
\newcommand{\bPsi}{\bs{\Psi}}
\newcommand{\bC}{\bs{C}}
\newcommand{\bR}{\bs{R}}
\newcommand{\bV}{\bs{V}}
\newcommand{\U}{U}
\newcommand{\Uop}{U_\op}
\newcommand{\Ucl}{U_\cl}
\newcommand{\p}{\hat{p}}
\newcommand{\hbp}{\hat{\bs{p}}}
\newcommand{\calH}{{\mathcal{H}}}
\newcommand{\calO}{\mathcal{O}}
\newcommand{\hO}{\hat{\calO}}
\newcommand{\calG}{\mathcal{G}}
\newcommand{\calP}{\mathcal{P}}
\newcommand{\Psit}{\widetilde{\Psi}}
\newcommand{\calGop}{\mathcal{G}_\op}
\newcommand{\calGcl}{\mathcal{G}_\cl}
\newcommand{\calV}{\mathcal{V}}
\newcommand{\calW}{\mathcal{U}}
\newcommand{\calU}{\mathcal{U}}
\newcommand{\calVsf}{\mathcal{V_\rm{sp}}}
\newcommand{\calL}{\mathcal{L}}
\newcommand{\IPair}{I^\rm{Pair}}
\newcommand{\dagg}{^{\dag}}
\newcommand{\intd}[1]{\int\frac{\dif #1}{2\pi}}
\newcommand{\half}{{\displaystyle\frac{1}{2}}}
\newcommand{\halfi}{{\displaystyle\frac{\im}{2}}}
\newcommand{\dint}{\int\!\!\!\int}
\newcommand{\tint}{\int\!\!\!\int\!\!\!\int}
\newcommand{\sint}{\tint\!\!\!\tint}
\newcommand{\ddint}{\dint\!\!\!\dint}
\newcommand{\dintd}[2]{\dint\frac{\dif #1}{2\pi}\,\frac{\dif #2}{2\pi}}
\newcommand{\tintd}[3]{\dint\frac{\dif #1}{2\pi}\,\frac{\dif #2}{2\pi}\,\frac{\dif #3}{2\pi}}
\newcommand{\ddintd}[4]{\ddint\frac{\dif #1}{2\pi}\,\frac{\dif #2}{2\pi}\,
\frac{\dif #3}{2\pi}\,\frac{\dif #4}{2\pi}}
\newcommand{\dddintd}[6]{\sint\frac{\dif #1}{2\pi}\,\frac{\dif #2}{2\pi}\,
\frac{\dif #3}{2\pi}\,\frac{\dif #4}{2\pi}\,\frac{\dif
#5}{2\pi}\,\frac{\dif #6}{2\pi}}
\newcommand{\ddt}{\frac{\partial}{\partial t}}
\newcommand{\intbr}{\int\dif\br\,}
\newcommand{\gamlim}{\ensuremath{\gamma\rightarrow 0}}
\newcommand{\vsp}{\vspace{0.5cm}}
\newcommand{\vvsp}{\vspace{0.25cm}}
\newcommand{\vvvsp}{\vspace{0.125cm}}
\newcommand{\mvsp}{\vspace{-0.5cm}}
\newcommand{\mvvsp}{\vspace{-0.25cm}}
\newcommand{\mvvvsp}{\vspace{-0.25cm}}
\newcommand{\mmvsp}{\vspace{-0.25cm}}
\newcommand{\mVsp}{\vspace{-1cm}}
\newcommand{\mVSP}{\vspace{-2cm}}
\newcommand{\Vsp}{\vspace{1cm}}
\newcommand{\Wsp}{\vspace{2cm}}
\newcommand{\hsp}{\hspace{0.5cm}}
\newcommand{\hhsp}{\hspace{0.25cm}}
\newcommand{\Hsp}{\hspace{1cm}}
\newcommand{\HSP}{\hspace{2cm}}
\newcommand{\mmhsp}{\hspace{-0.25cm}}
\newcommand{\mhsp}{\hspace{-0.5cm}}
\newcommand{\mHsp}{\hspace{-1cm}}
\newcommand{\mHSP}{\hspace{-2cm}}
\newcommand{\ö}{\"{o}}
\newcommand{\Ö}{\"{O}}
\newcommand{\ä}{\"a}
\newcommand{\å}{\aa}
\newcommand{\Å}{\AA}
\newcommand{\ue}{\"{u}}
\newcommand{\hpsi}{\hat{\psi}}
\newcommand{\wt}[1]{\widetilde{#1}}
\newcommand{\WU}{\widetilde{U}}
\renewcommand{\it}{\textit}
\renewcommand{\bf}{\textbf}
\newcommand{\bfit}[1]{\textbf{\it{#1}}}
\newcommand{\bful}[1]{\textbf{\ul{#1}}}
\newcommand{\ul}{\underline}
\newcommand{\itul}[1]{\it{\ul{#1}}}
\newcommand{\abs}[1]{|{#1}|}
\newcommand{\Abs}[1]{\big|{#1}\big|}
\newcommand{\eq}{\eqref}
\newcommand{\rarr}{\rightarrow}
\newcommand{\larr}{\leftarrow}
\newcommand{\lrarr}{\leftrightarrow}
\newcommand{\LRarr}{\Longleftrightarrow}
\newcommand{\Rarr}{\Rightarrow}
\newcommand{\Lrarr}{\Longrightarrow}
\newcommand{\rr}{r_{12}}
\newcommand{\frr}{\frac{1}{r_{12}}}
\newcommand{\brr}{\br_{12}}
\newcommand{\DF}{D_{\rm{F}\nu\mu}}
\newcommand{\DFS}{D_{\rm{F}}}
\newcommand{\DFmn}{D_{\rm{F}\mu\nu}}
\newcommand{\DFCij}{D^C_{\rm{Fij}}}
\newcommand{\SF}{S_{\rm{F}}}
\newcommand{\hSF}{\hat{S}_{\rm{F}}}
\newcommand{\hh}{\hat{h}}
\newcommand{\QED}{\rm{QED}}
\newcommand{\CQED}{\rm{CQED}}
\newcommand{\DQED}{\rm{DQED}}
\newcommand{\ren}{\rm{ren}}
\newcommand{\bou}{\rm{bou}}
\newcommand{\free}{\rm{free}}
\renewcommand{\sp}[2]{\bra{#1}{#2}\rangle}
\newcommand{\SP}[2]{\Bigbra{#1}{#2}\Big\rangle}
\newcommand{\qand}{\quad\rm{and}\quad}
\newcommand{\V}{V}
\newcommand{\VC}{V_\rm{C}}
\newcommand{\VF}{V_\rm{F}}
\newcommand{\VG}{V_\rm{G}}
\newcommand{\VspC}{V_\rm{TC}}
\newcommand{\Vsr}{V_\rm{sr}}
\newcommand{\MC}{\calM_\rm{C}}
\newcommand{\IC}{I^\rm{C}}
\newcommand{\ICC}{I^\rm{C}_\rm{C}}
\newcommand{\ICT}{I^\rm{C}_\rm{T}}
\newcommand{\UT}{U_\rm{T}}
\newcommand{\VT}{V_\rm{T}}
\newcommand{\vT}{v_\rm{T}}
\newcommand{\HD}{H_\rm{D}}
\newcommand{\fC}{f^\rm{C}}
\newcommand{\fCC}{f^\rm{C}_\rm{C}}
\newcommand{\fCT}{f^\rm{C}_\rm{T}}
\newcommand{\fF}{f^\rm{F}}
\newcommand{\VB}{V_\rm{B}}
\newcommand{\Vsf}{V_\rm{sp}}
\newcommand{\Vgsf}{\calV_\rm{T}}
\newcommand{\Vsfp}{V_\rm{sp}}
\newcommand{\Usfp}{U_\rm{sp}}
\newcommand{\Msfp}{\calM_\rm{sp}}
\newcommand{\MVx}{\calM_\rm{Vx}}
\newcommand{\VSE}{V_\rm{SE}}
\newcommand{\VVert}{V_\rm{Vx}}
\newcommand{\VVx}{V_\rm{Vx}}
\newcommand{\VspGen}{V_\rm{tr}^\rm{Gen}}
\newcommand{\VTC}{V_\rm{TC}}
\newcommand{\calUsp}{\calU_\rm{sp}}
\newcommand{\Usf}{U_\rm{sp}}
\newcommand{\Ssf}{S_\rm{T}}
\newcommand{\Msf}{\calM_\rm{sp}}
\newcommand{\MT}{\calM_\rm{T}}
\newcommand{\Ksf}{{\cal K_\rm{sp}}}
\newcommand{\MSE}{\calM_\rm{SE}}
\newcommand{\SSE}{S_\rm{SE}}
\newcommand{\USE}{U_\rm{SE}}
\newcommand{\Coul}{\rm{Coul}}
\newcommand{\calK}{\mathcal{\kappa}}
\newcommand{\calKc}{I_c}
\newcommand{\calF}{\mathcal{F}}
\newcommand{\calM}{{\mathcal{M}}}
\newcommand{\calR}{\mathcal{\hat{R}}}
\newcommand{\calB}{\mathcal{B}}
\newcommand{\GQ}{\Gamma_Q}
\newcommand{\GV}{\Gamma V}
\newcommand{\Gam}{\Gamma}
\newcommand{\GQV}{\GQ V}
\newcommand{\GI}{\calG^\rm{I}}
\newcommand{\bGQ}{\bs{\Gamma_Q}}
\newcommand{\Gk}{\bs{\Gamma_Q}}
\newcommand{\Gv}{\Gamma_Q}
\newcommand{\TD}{T_\rm{D}}
\newcommand{\F}{\rm{F}}
\renewcommand{\P}{\bs{P}}
\newcommand{\Util}{\widetilde{U}}
\newcommand{\Ucov}{U_\Cov}
\renewcommand{\S}{S}
\newcommand{\Vtil}{\widetilde{V}}
\newcommand{\Om}{\Omega}
\newcommand{\Omsf}{\Omega_\rm{sp}}
\newcommand{\Omi}{\Om_\rm{I}}
\newcommand{\OmI}{{\Om_\rm{I}}}
\newcommand{\om}{\omega}
\newcommand{\OM}{\mathcal{\boldsymbol{\Om}}}
\newcommand{\Ombar}{\bar{\Omega}}
\newcommand{\Vbar}{\bar{V}}
\newcommand{\VI}{V_{12}}
\newcommand{\VR}{V_{\rm{R}}}
\newcommand{\VRbar}{\Bar{V}_{\rm{R}}}
\newcommand{\calE}{{\mathcal{E}}}
\newcommand{\CALE}{{\mathcal{E}}}
\newcommand{\la}{\lambda}
\newcommand{\FLL}{F_\rm{LL}}
\newcommand{\ka}{|\bs{k}|}
\newcommand{\norm}[1]{||{#1}||}
\newcommand{\Norm}[1]{\big|\big|{#1}\big|\big|}
\newcommand{\NORM}[1]{\Big|\Big|{#1}\Big|\Big|}
\newcommand{\Fr}{Fr\'echet }
\newcommand{\Ga}{G\^ateaux }
\newcommand{\epsn}{\ensuremath{\epsilon}}
\newcommand{\Der}[1]{\frac{\dif}{\dif#1}}
\newcommand{\der}[2]{\frac{\dif#1}{\dif#2}}
\newcommand{\ave}[1]{\langle #1\rangle}
\newcommand{\Ave}[1]{\Big\langle #1\Big\rangle}
\newcommand{\BB}[2]{\Big\{{#1}\,\Big|\:{#2}\Big\}}
\newcommand{\LL}{L^1\cap L^3}
\newcommand{\partdelta}[2]{\frac{\delta #1}{\delta #2}}
\newcommand{\pd}[1]{\frac{\delta #1}{\delta \calE}}
\newcommand{\pda}[1]{\frac{\delta^* #1}{\delta \calE}}
\newcommand{\Pda}[1]{\frac{\delta^*}{\delta #1}}
\newcommand{\Partdelta}[1]{\frac{\delta }{\delta #1}}
\newcommand{\Pd}[1]{\frac{\delta }{\delta #1}}
\newcommand{\partdeltan}[3]{\frac{\delta^#1 #2}{\delta #3^#1}}
\newcommand{\partdeltanp}[3]{\frac{\delta^{(#1)} #2}{\delta
#3^{(#1)}}}
\newcommand{\pdna}[2]{\frac{\delta^{^*#1} #2}{\delta\calE^#1}}
\newcommand{\pdn}[2]{\frac{\delta^{#1} #2}{\delta\calE^{#1}}}
\newcommand{\pdnp}[2]{\frac{\delta^{(#1)} #2}{\delta\calE^{(#1)}}}
 \newcommand{\ett}{^{(1)}}
\newcommand{\nol}{^{(0)}}
\newcommand{\tva}{^{(2)}}
\newcommand{\tre}{^{(3)}}
\newcommand{\fyr}{^{(4)}}
\newcommand{\enh}{^{(1/2)}}
\newcommand{\treh}{^{(3/2)}}
\newcommand{\femh}{^{(5/2)}}
\newcommand{\n}{^{(n)}}
\newcommand{\m}{^{(m)}}
\newcommand{\nm}{^{(n-m)}}
\newcommand{\nett}{^{(n-1)}}
\newcommand{\ntva}{^{(n-2)}}
\newcommand{\PE}{P_\mathcal{E}}
\newcommand{\GE}{\Gamma(\calE)}
\newcommand{\GEE}{\Gamma(E)}
\newcommand{\GQEE}{\Gamma_Q(E)}
\newcommand{\GQE}{\Gamma_Q(\calE)}
\newcommand{\GQEP}{\Gamma_Q(\EP)}
\newcommand{\GEN}{\Gamma(E_0)}
\newcommand{\GQEN}{\Gamma_Q(E_0)}
\newcommand{\GEP}{\Gamma(\calE')}
\newcommand{\GaV}{\Gamma V}
\newcommand{\GaVg}{\GQ V}
\newcommand{\PEP}{P_{\calE'}}
\newcommand{\QE}{Q_{\calE}}
\newcommand{\QEP}{Q_{\calE'}}
\newcommand{\PEPP}{P_{\calE''}}
\newcommand{\EP}{\calE'}
 \newcommand{\limgam}{\lim_{\gamlim}}
 \newcommand{\Ugam}[1]{U_\gamma(#1,-\infty)}
\newcommand{\Ugamtil}[1]{\widetilde{U}_\gamma(#1,-\infty)}
\newcommand{\Ugamt}{\widetilde{U}_\gamma}
\newcommand{\img}{\im\gamma}
\newcommand{\ime}{\im\eta}
\newcommand{\pbar}{\not\!p}
\newcommand{\psl}{\!\not\!p\,}
\newcommand{\Asl}{\not\!\! A\,}
\newcommand{\qsl}{\not\!q}
\newcommand{\ksl}{\not\!k}
\newcommand{\lsl}{\not\!l}
\newcommand{\nott}{\not\!\!}
\newcommand{\wtp}{\widetilde{p}}
\newcommand{\wtq}{\widetilde{q}}
\newcommand{\wtk}{\widetilde{k}}
\newcommand{\Deltag}{\Delta_\gamma}
\newcommand{\Deltatg}{\Delta_{2\gamma}}
\newcommand{\Deltafg}{\Delta_{4\gamma}}
\renewcommand{\bar}{\setlength{\unitlength}{0.6cm}\put(0,0.6){\line(1,0){0.4}}}
\newcommand{\fbar}{\setlength{\unitlength}{0.6cm}\put(0,0.4){\line(1,0){0.4}}}
\newcommand{\MSC}{\rm{MSC}}
\newcommand{\IF}{\rm{IF}}
                                                  \newcommand{\Htil}{\widetilde{H}}
                                                  \newcommand{\Ubar}{\bar{U}}
                                                  \newcommand{\Hbar}{\bar{V}}
                                                  \newcommand{\Udot}{\dot{U}}
                                                  \newcommand{\Ubardot}{\dot{\Ubar}}
                                                  \newcommand{\Utildot}{\dot{\Util}}
                                                  \newcommand{\Cdot}{\dot{C}}
                                                  \newcommand{\Ombardot}{\dot{\Ombar}}
                                                  \newcommand{\bsdot}{\bs{\cdot}}

\newcommand{\npartdelta}[3]{\frac{\delta^#1 #2}{\delta #3^#1}}
\newcommand{\ip}[1]{| #1 \rangle \langle #1 |}
\newcommand{\con}{\mathrm{con}}
\newcommand{\ph}{\mathrm{ph}}
\newcommand{\bA}{\bs{A}}
\newcommand{\Pmu}{\partial^\mu}
\newcommand{\Pnu}{\partial^\nu}
\newcommand{\pmu}{\partial_\mu}
\newcommand{\pnu}{\partial_\nu}
\newcommand{\bAT}{\bs{A}_\perp}
\newcommand{\bAL}{\bs{A}_\parallel}
\newcommand{\bET}{\bs{E}_\perp}
\newcommand{\bEL}{\bs{E}_\parallel}
\newcommand{\bB}{\bs{\rm{B}}}
\newcommand{\bE}{\bs{\rm{E}}}
\newcommand{\bn}{\bs{n}}
\newcommand{\beps}{\bs{\eps}}
\newcommand{\HI}{H_\rm{int,I}}
\newcommand{\calHI}{{\cal H}_\rm{int,I}}
\newcommand{\ih}{\frac{\im}{\hbar}}
\newcommand{\bkx}{\bk\bdot\bx}
\newcommand{\halfS}{{\textstyle\frac{1}{2}\,}}
\newcommand{\h}{\hat{h}}
\newcommand{\DFnu}{D_{\rm{F}\nu\mu}}
\newcommand{\Eab}{E_{ab}}
\newcommand{\rE}{{\red E}}
\newcommand{\sphline}{\hline\vsp}
\newcommand{\Ram}[4]
{\begin{picture}(0,0)(0,0)\setlength{\unitlength}{1cm}
    \put(#1,#2){\Ebox{#3}{#4}}
  \end{picture}}
                                                  \newcommand{\Ers}{E_{rs}}
                                                  \newcommand{\Etu}{E_{tu}}
                                                  \newcommand{\Eru}{E_{ru}}
                                                  \newcommand{\Ecd}{E_{cd}}
                                                  \newcommand{\Erd}{E_{rd}}
                                                  \newcommand{\epsi}{\epsilon}

\setlength{\unitlength}{0.75cm}

\title{Dimensional regularization of the free-electron self-energy
and vertex correction in Coulomb gauge}
\author{Ingvar Lindgren\\ Physics Department, University of Gothenburg,
G\öteborg, Sweden}
 %\pacs{}
\date{\today} \maketitle

\abstract There is presently a great interest in studying static
and dynamic properties of highly charged ions that can be produced
in large particle accelerators, like that at GSI in Darmstadt. To
perform corresponding theoretical calculations with great accuracy
requires a highly developed machinery of computational methods
that have not until recently been available. In order to combine
many-body perturbation theory with quantum electrodynamics, the
calculations have generally to be performed in the Coulomb gauge,
where applications have not been so developed as in, for instance,
the Feynman gauge. Formulas for the free-electron self energy and
vertex correction have been given without derivation by Adkins
(Phys. Rev. D27, 1814 (1983); Phys. Rev. D34, 2489 (1986)). In the
present paper the formulas of Adkins are verified with detailed
derivations.

\newcommand{\pdk}{\bp\bsdot\bk}
\newcommand{\ppdk}{\bp'\bsdot\bk}
\newcommand{\pdpp}{\bp\bsdot\bp'}
\newcommand{\pdq}{\bp\bsdot\bq}
\newcommand{\ppdq}{\bp'\bsdot\bq}

\section{Introduction}
There is now an increased interest in studying the effects of
quantum-electrodynamics in combination with electron correlation
in electronic systems, particularly in connection with experiments
on highly charged ions (see ref.~\cite{ILBook11} and references
therein). In order to take full advantage of the development in
atomic many-body theory~\cite{LM86}, it is necessary to perform
the calculations in the Coulomb gauge. Dimensional regularization
in that gauge is more complicated than in, for instance, the
Feynman gauge and has to date to our knowledge not been used in
practical calculations. Such calculations have now been performed
at our department~\cite{DanJoh11}, and as a background we have
reconsidered the formulas derived by Adkins some time
ago~\cite{Adkins83,Adkins86}. Adkins gives only the final results
without any derivation, and we have found that it might be useful
to produce full derivations of the formulas. One derivation is
reproduced here, and an alternative treatment is being published
separately~\cite{JohDan11}.

In the book cited above~\cite{ILBook11} the dimensional
regularization of the free-electron self-energy and vertex
correction are treated in the Feynman gauge and the self-energy
also in the Coulomb gauge, while the vertex correction in the
latter gauge was found to be too complex to include in the book.
For that reason it is instead reproduced here.

\section{Free-electron self energy in Coulomb gauge}

We shall mainly follow Adkins~\cite{Adkins83} in regularizing the
free-electron self energy in the Coulomb gauge.\footnote{This part
is essentially reproduced from the above-mentioned
book~\cite[sect. 12.5]{ILBook11} with due permission from the
publisher.} We start from the expressions for the self-energy
($c=1$)
\begin{equation}\label{SEB}
   \Sigma^\rm{free}(p)=\im e^2\intDD{k}\;\gamma^\nu
   \frac{\psl-\ksl+m}{(p-k)^2-m^2+\ime}\,\gamma^\mu \DF(k)
\end{equation}
and for the photon propagator
\begin{equation}   \label{DC1}
  D_{\rm{F}\mu\nu}^{\rm{C}}(k;\bk)=\frac{1}{\epsi_0}\,\Bigg[
  \frac{\delta_{\mu,0}\delta_{\nu,0}}{\bk^2}-
  \delta_{\mu,i}\delta_{\nu,j}\Big(g_{ij}+\frac{k_i k_j}{\bk^2}\Big)
  \frac {1}{k^2+\ime}\Bigg]
\end{equation}
The three terms in the propagator correspond to the \it{Coulomb,
Gaunt and scalar-retardation} parts.

\subsection{Coulomb contribution}
The \bfit{Coulomb part} of the self energy becomes
\begin{eqnarray}\label{SECoul}
   &&\frac{\im e^2}{\epsi_0}\intDD{k}\;
   \frac{\gn(\psl-\ksl+m)\gn}{(p-k)^2-m^2+\ime}\,\frac{1}{\bk^2+\ime}\\
   &=&\frac{\im e^2}{\epsi_0}\intDD{k}\;
   \frac{\wtp-\wtk+m}{(p-k)^2-m^2+\ime}\,\frac{1}{\bk^2+\ime}
\end{eqnarray}
using the commutation rules in Appendix \ref{sec:gamma} (Eq.
\ref{gammaRel}). With $q=-p\,$ and $s=p^2-m^2$ the denominator is
of the form $k^2+2kq+s$ and we can apply the formulas \eqref{DI4k}
and \eqref{DI5k} in Appendix \ref{sec:IntReg} for dimensional
regularization with $D=4-\epsi$ ($n=1$). This gives with $k^0\rarr
-q^0=p^0$, $k_i\rarr -q_iy=p_iy$, $\bgk=-\giki\rarr\bgp y$ and
$w=p^2y^2+(1-y)yp_0^2-(p^2-m^2)y$
\begin{eqnarray*}
   &&\frac{\im e^2}{\epsi_0}\intDim{k}\,
   \frac{\wtp\,-\wtk+m}{k^2+2kq+s+\ime}   \,\frac{1}{\bk^2+\ime}\nn
   &=&(\im)^2(-1)^n\frac{e^2}{\epsi_0\,(4\pi)^{D/2}}\,\int_0^1\frac{\dif y}{\sqrt y}\;
    \big[\bgp(1-y)+m\big]\,\frac{\Gamma(\epsi/2)}{w^{\epsi/2}}\;\nn
\end{eqnarray*}
The Gamma function can be expanded as
  \[\Gamma(\epsi/2)=\frac{2}{\epsi}-\gamma_\rm{E}+\cdots\]
where $\gamma_\rm{E}=0.5722...$  is Euler's constant, and
furthermore
  \[\Big(\frac{m^2}{w}\Big)^{\epsi/2}
  =1-\frac{\epsi}{2}\,\ln\Big(\frac{w}{m^2}\Big)+\cdots\]
  \[\frac{1}{(4\pi)^{D/2}}=\frac{1}{(4\pi)^2}\Big(1+\frac{\epsi}{2}\ln4\pi+\cdots\Big)\]
This yields
\begin{eqnarray}\label{Dlim}
&&\frac{\Gamma(\epsi/2)}{(4\pi)^{D/2}}\,\Big(\frac{m^2}{w}\Big)^{\epsi/2}=\nn
&&\mhsp\frac{1}{(4\pi)^2}\Big(2/\epsi-\gamma_\rm{E}+\cdots\Big)\,\Big(1+\frac{\epsi}{2}\ln4\pi+\cdots\Big)
   \Big(1-\frac{\epsi}{2}\,\ln \big(w/m^2\big)+\cdots\Big)\nn
   &=&\frac{1}{(4\pi)^2}\Big[\Delta-\ln
   \Big(\frac{w}{m^2}+\cdots\Big)\Big]
\end{eqnarray}
where
\begin{equation}\label{DeltaDim}
  \boxed{\Delta=\frac{2}{\epsi}-\gamma_\rm{E}+\ln4\pi+\cdots}
\end{equation}
This leads to the Coulomb contribution \eqref{SECoul}, omitting
the factor $m^{-\epsi/2}$,
\begin{eqnarray*}
   K\int_0^1\frac{\dif y}{\sqrt y}\;\Big(\bgp\,(1-y)+m\Big)
   \Big(\Delta-\ln(yX)\Big)
\end{eqnarray*}
where in relativistic
units\,\footnote{$c=\epsi_0=1,\,e^2=4\pi\alpha$ ($\alpha$
fine-structure constant)}
\begin{equation}\label{K}
K=\frac{e^2}{\epsi_0\,(4\pi)^2}=\frac{\alpha}{4\pi}
\end{equation}
and $w=m^2y\,X$, $X=1+(\bp^2/m^2)(1-y)$. This leads to
\begin{eqnarray*}
   &&K\int_0^1\frac{\dif y}{\sqrt y}\;
   \Big((\bgp\,(1-y)+mc\Big)\Big(\Delta -\ln y-\ln X\Big)
\end{eqnarray*}
and the Coulomb part \eqref{SECoul} of the free-electron self
energy becomes (times K)
\begin{eqnarray}\label{SECoul1}
   \mhsp\boxed{\Big(\frac{4}{3}\bgp +2m\Big)\Delta +\Big(\frac{32}{9}\bgp+4m\Big)
   -\int_0^1\frac{\dif y}{\sqrt y}\;\Big((\bgp\,(1-y)+m\Big)\ln
   X}\Hsp\hsp
\end{eqnarray}

\subsection{Gaunt contribution}
The \bfit{Gaunt term} becomes, using \eqqref{SEB} and the second
term of \eqqref{DC1},
\begin{eqnarray}\label{SEGaunt}
   &&-\frac{\im e^2}{\epsi_0}\intDD{k}\;
   \frac{\gii(\psl\,-\ksl+m)\gi}{(p-k)^2-m^2+\ime}\,\frac{1}{k^2+\ime}
\end{eqnarray}
Using the Feynman integral \eqref{FI2} (second version) in
Appendix \ref{sec:FI} with $a=k^2$ and
\makebox{$b=(p-k)^2-m^2c^2$}, this can be expressed\,\footnote{We
use the convention that $\mu,\,\nu,..$ represent all four
components (0,1,2,3), while $i,\,j,..$ represent the vector part
(1,2,3).}
\begin{eqnarray}\label{SEGaunt1}
   &&-\frac{\im e^2}{\epsi_0}\int_0^1\dif x \intDD{k}\;
   \frac{\gamma_i\big(\psl\,\!-\ksl+m\big)\gamma^i}{\big[k^2+(p^2-2pk-m^2)x\big]^2}\nn
   &&=-\frac{\im e^2}{\epsi_0}\int_0^1\dif x\intDD{k}\;
   \frac{(3-\epsi)m-(2-\epsi)(\psl\,-\ksl)-\wtp+\wtk}{\big[k^2+(p^2-2pk-m^2)x\big]^2}
\end{eqnarray}
after applying the commutation rules in Eq. \ref{gammaDim}.

With the substitutions $k\rarr-q=px$ and $s=(p^2-m^2)x$ we can
apply the equations (\ref{DI4} and \ref{DI5}) in Appendix
\ref{sec:IntReg}, leading to
\begin{eqnarray*}
   &&-\frac{\im e^2}{\epsi_0}\int_0^1\dif x\;\intDim{k}\;
   \frac{(3-\epsi)m-(2-\epsi)(\psl\,-\ksl)-\wtp+\wtk}{\big[k^2+2kq+s)\big]^2}\nn
   &=&\frac{e^2}{\epsi_0(4\pi)^{D/2}}\int_0^1\dif x\;
   \Big[(3-\epsi)m-(2-\epsi)\psl(1-x)-\wtp(1-x)\Big]
    \frac{\Gamma(\epsi/2)}{w^{\epsi/2}}\;\nn
   &=&\frac{e^2}{\epsi_0(4\pi)^{D/2}}\int_0^1\dif x\;
   \Big[-(1-x)\big(3\gnp-\bgp\big)+3m
   +\epsi\big((1-x)\psl\,-m\big)\Big]\frac{\Gamma(\epsi/2)}{w^{\epsi/2}}\;
\end{eqnarray*}
where  $w=q^2-s=p^2x^2-(p^2-m^2)x=m^2xY$. This yields the Gaunt
contribution \eqref{SEGaunt} (times K)
\begin{eqnarray*}
   &&\mhsp -\int_0^1\dif x\;\Bigg\{\Big[(1-x)\big(3\gnp-\bgp\big)-3m\Big]
   \Big[\Delta-\ln (xY)\Big]-2\big((1-x)\psl\,-m\big)\Bigg\}
\end{eqnarray*}
using the relation \eqref{Dlim} and the fact that
$\;\epsi\,\Delta\rarr2\;$ as $\;\epsi\rarr0$. Then the Gaunt part
of the free-electron self energy becomes\,\footnote{Note misprint
in first bracket of Eq. (12.113) of ref. \cite{ILBook11}} \small
\begin{eqnarray}\label{SEGaunt2}
   \mHsp\boxed{\Big[-\half\,\big(3\gnp-\bgp\big)+3m\Big]\Delta
   -\frac{5}{4}\,\gnp-\frac{1}{4}\,\bgp+m
   +\int_0^1\dif x\;\Big[(1-x)\big(3\gnp-\bgp\big)-3m\Big]\ln Y}\mHsp\nn
\end{eqnarray}\normalsize

\subsection{Scalar-retardation contribution}
Finally, the \bfit{scalar-retardation part} becomes similarly,
using the third term of \eqqref{DC1} and the commutation rules
\eqref{gammaAC},
\begin{eqnarray}\label{SEScalRet}
   &-&\frac{\im e^2}{\epsi_0}\intDD{k}\;\frac{\gi k_i\,(\psl\,-\ksl+m)\,\gamma^j k_j}
   {(p-k)^2-m^2+\ime}\,\frac{1}{\bk^2}\,\frac{1}{k^2+\ime}\nn
   &=&\frac{\im e^2}{\epsi_0}\intDD{k}\;\frac{\gamma^i k_i\gamma^j k_j(\psl\,-\ksl-m)
   -2\gamma^ik_i(k^jp_j-k^jk_j)}{(p-k)^2-m^2+\ime}\,\frac{1}{\bk^2}\,\frac{1}{k^2+\ime}\nn
   &=&-\frac{\im e^2}{\epsi_0}\intDD{k}\;
   \frac{\psl\,-\wtk-m+2\gamma^ik_i\,k^jp_j/\bk^2}{(p-k)^2-m^2+\ime}\,\frac{1}{k^2+\ime}
\end{eqnarray}
with $\gamma^i k_i\gamma^j k_j=-\bk^2=-k_ ik_i$. With the same
substitutions as in the Gaunt case this becomes
\begin{eqnarray}\label{SEScalRet2}
  -\frac{\im e^2}{\epsi_0}\int_0^1\dif x\,\intDim{k}\;
  \frac{\psl\,-\wtk-m+2\gamma^ik_i\,k^jp_j/\bk^2}
{\big[k^2-2pkx+(p^2-m^2)x\big]^2}
\end{eqnarray}
With the substitutions $k\rarr-q=px$ and $s=(p^2-m^2)x$ the first
part is of the form \eqqref{DI4} and \eqqref{DI5} in Appendix
\ref{sec:IntReg} and becomes
\begin{eqnarray}\label{SEScalRet3}
  \frac{e^2c}{\epsi_0(4\pi)^{D/2}}\int_0^1\dif x\,
  \big[\psl\,-\wtp\,x-m\big]\;\frac{\Gamma(\epsi/2)}{w^{\epsi/2}}
\end{eqnarray}
and with \eqqref{Dlim}
\begin{eqnarray}\label{SEScalRet4}
   K\int_0^1\dif x \big[\psl\,-\wtp\,x-m\big]\,\Big(\Delta-\ln (xY)\Big)
\end{eqnarray}
with $K=e^2/(\epsi_0\,(4\pi)^2)$ and $w$ being the same as in the
Gaunt case, $w=q^2-s=p^2x^2-p^2x+m^2x=m^2xY$.

The second part of \eqqref{SEScalRet2} is of the form
\eqqref{DI6k} and becomes ($k_ik^j\rarr
q_i\,q^j\,y^2=p_i\,p^j\,x^2y^2$ in first term, $\rarr -\halfS
g^i_j=-\halfS \delta_{ij}$ in second)
\begin{eqnarray*}
   &&K\int_0^1\dif x\int_0^1\dif y\,\sqrt y\;\Bigg\{2\gamma^ip_i\,p^i\,p_j\,p^j\,
   \frac{\Gamma(1+\epsi/2)}{w^{1+\epsi/2}}
   -\gamma^jp_j\;\frac{\Gamma(\epsi/2)}{w^{\epsi/2}}\Bigg\}\nn
   &&K\int_0^1\dif x\int_0^1\dif y\,\sqrt y\;\Bigg\{\frac{2\gamma^ip_i\,p^jp_j}{m^2}\,
   \frac{xy}{Z}-\gamma^jp_j\Big(\Delta-\ln (xyZ)\Big)\Bigg\}\nn
   &=&K\int_0^1\dif x\int_0^1\dif y\,\sqrt y\;\Bigg\{\frac{2\bgp\,\bp^2}{m^2}\,
   \frac{xy}{Z}+\bgp\Big(\Delta-\ln (xyZ)\Big)\Bigg\}
\end{eqnarray*}
with
$w=xy\big[-\bp^2xy+p_0^2x-p^2+m^2\big]=xy\big[\bp^2(1-xy)-p_0^2(1-x)+m^2\big]=m^2Zxy$

Integration by parts of the first term yields (times $K$), noting
that $dZ/dy=-\bp^2x$,
  \[-\int_0^1\dif x\,\Big[\sqrt y\,y\,2\bgp\ln Z\Big]_0^1+
  3\int_0^1\dif x\int_0^1\dif y\,\sqrt y\;\bgp\,\ln Z\]
The total scalar-retardation part then becomes (with $Z(y=1)=Y$)
\begin{eqnarray*}
  &&\int_0^1\dif x \big[\psl\,-\wtp\,x-m\big]\,\Big(\Delta-\ln (xY)\Big)
  -\int_0^1\dif x\,2\bgp\ln Y\nn
  &+&3\int_0^1\dif x\int_0^1\dif y\,\sqrt y\;\bgp\,\ln Z
  +\int_0^1\dif x\int_0^1\dif y\,\sqrt y\;\bgp\Big(\Delta-\ln (xyZ)\Big)
\end{eqnarray*}
or
\begin{eqnarray*}
  &&\int_0^1\dif x \Big(\gnp(1-x)-\bgp(1+x)-m\Big)\,\Big(\Delta-\ln x\Big)
  +\int_0^1\dif y\,\sqrt y\;\bgp\,\Delta\nn
  &-&\int_0^1\dif x \Big(\gnp(1-x)  -\bgp(1-x)-m\Big)\,\ln Y\nn
  &-&\int_0^1\dif x\int_0^1\dif y\,\sqrt y\;\bgp\,\ln (xy)
  +2\int_0^1\dif x\int_0^1\dif y\,\sqrt y\;\bgp\,\ln (xy)\nn
  &-&3\int_0^1\dif x\int_0^1\dif y\,\sqrt y\;\bgp\,\ln Z
\end{eqnarray*}
Then the scaler-retardation part of the free-electron self energy
becomes
\begin{eqnarray*}
  &&\Big(\frac{1}{2}\,\gnp-\frac{5}{6}\,\bgp-m\Big)\Delta
  +\frac{3}{4}\,\gnp-\frac{5}{36}\,\bgp-m\nn
  &&\mHsp-\int_0^1\dif x \Big(\gnp(1-x)  -\bgp(1-x)-m\Big)\,\ln Y
  -3\int_0^1\dif x\int_0^1\dif y\,\sqrt y\;\bgp\,\ln Z
\end{eqnarray*}

Summarizing all contributions yields \noindent\bfit{\;the
mass-renormalized free-electron self energy in the Coulomb gauge}
\begin{eqnarray}\label{SETot} &&\mHsp\mhsp\frac{e^2}{\epsi_0\,(4\pi)^2}
  \Bigg[-\Big(\psl-m\Big)\Delta-\half\,\gnp+\frac{19}{6}\,\bgp
  -\int_0^1\frac{\dif y}{\sqrt y}\;\Big(\bgp\,(1-y)+m\Big)\ln X\nn
  &+&2\int_0^1\dif x\;\big[(1-x)\psl-m\big]\ln Y
  +\int_0^1\dif x\int_0^1\dif y\,\sqrt y\;2\bgp \,\ln
  Z\Bigg]%\Ram{-10.25}{-0.5}{13}{2.45}\nn
\end{eqnarray}
where we have subtracted the on-shell ($\psl=m$) value,
$Km(3\Delta+4)$ (which is the same as in the Feynman
gauge~\cite[Eq. (12.103)]{ILBook11}). The expressions for $X, Y,
Z$ are given in the text. The result is in agreement with that of
Adkins~\cite{Adkins83}.

\normalsize \section{Free-electron vertex function in Coulomb
gauge}
\newcommand{\ps}{p^\sigma}
\newcommand{\ks}{k^\sigma}
\newcommand{\pps}{p^{'\sigma}}
\newcommand{\ppmu}{p^{'\mu}}
\newcommand{\ppn}{p^{'0}}
\newcommand{\dsl}{\!\not\!d\,}
\newcommand{\pn}{p^0}
\newcommand{\kn}{k^0}

Next, we consider the free-electron vertex function in the Coulomb
gauge and start from the expression~\cite[Eq. (1)]{Adkins86}
\begin{eqnarray}\label{Vertex}
  \mhsp\Lambda_\sigma(p,p\,')&=&\im e^2\intDD{k}\;
  \gnu \frac{\psl\,'-\ksl+ m}{(p\,'-k)^2-m^2+\ime}\,
  \gs\frac{\psl-\ksl+m}{(p-k)^2-m^2+\ime}\,\gmu\DF\Hsp
\end{eqnarray}

We restrict ourselves here to the case, where $\gs=\gn$, which
implies that the interaction at the vertex is scalar and can be
the \it{Coulomb interaction with the nucleus or with another
electron}.

\newcommand{\bd}{\bs{d}}
\renewcommand{\bgd}{\bgamma\bsdot\bs{d}}
\newcommand{\pslp}{\psl\,'}
\newcommand{\bsd}{\bgamma\bsdot}

\subsection{Coulomb contribution}\label{sec:Coulomb}
Inserting the \bfit{Coulomb part} of the photon propagator
\eqref{DC1} into the vertex expression \eqref{Vertex}, yields
\begin{eqnarray*}
  &&\mHsp\frac{\im e^2}{\epsi_0}\intDD{k}\;
  \gn\frac{\psl\,'-\ksl+ m}{(p\,'-k)^2-m^2+\ime}\,
  \gn\frac{\psl-\ksl+m}{(p-k)^2-m^2+\ime}\,\gn\frac{1}{\bk^2}
\end{eqnarray*}
Using the relation $\psl\gn=\gn\wtp$ (see Appendix
\ref{sec:gamma}) and the Feynman parametrization \eqref{FI2} in
Appendix \ref{sec:FI}, this leads to
\begin{eqnarray*}
  \frac{\im e^2}{\epsi_0}\int_0^1\dif x
 \intDD{k}\,\frac{\gn(\psl\,'-\ksl+ m)(\wtp-\wtk+ m)}
  {\big[k^2-2pk+p^2-m^2-(p^2-p'^2-2kp+2kp')x\big]^2} \,\frac{1}{\bk^2}
\end{eqnarray*}
With \[q=-(1-x)p-p'x=-(p-dx)\,;\hsp d=p-p'\]
\[s=(1-x)p^2+p'^2x-m^2\] the denominator is of the form
$k^2+2kq+s$. Introducing the dimension $D=4-\epsi$, we then have
\begin{eqnarray*}
  &&\frac{\im e^2}{\epsi_0}\int_0^1\dif x
 \intDim{k} \frac{\gn(\psl\,'-\ksl+ m)(\wtp-\wtk+ m)}
 {(k^2+2pq+s)^2}\,\frac{1}{\bk^2}
\end{eqnarray*}
and we can then apply the formulas \eqqref{DI4k} to \eqqref{DI6k}
in Appendix \ref{sec:IntReg}. In \eqqref{DI5k} and the first part
of \eqqref{DI6k} we make the substitution $k^\mu\rarr -q^\mu
y-\delta_{\mu,0}\,q_0(1-y)$ or
  \[\bk\rarr-\bq y=\bp y-\bd xy\hsp k_0\rarr-q_0=p_0-d_0x\]
  \[\ksl\rarr\bgq\,y-\gn q_0=-(\bgp-\bgd x)y+\gn(p_0-d_0x)\]
  \[\wtk\rarr-\bgq\,y-\gn q_0=(\bgp-\bgd x)y+\gn(p_0-d_0x)\]
and in the second part of \eqqref{DI6k}
  \[k^\mu k^\nu\rarr-\halfS \big[g^{\mu\nu}+\delta_{\mu,0}
  \delta_{\nu,0}(1-y)/y\big]\]
  \[\ksl\;\wtk=k_0^2-\gi k_i\,\gj k_j\rarr\halfS(-1/y+\gi\gii)
  =\halfS(3-\epsi-1/y)\]
This yields in analogy with the self energy
\begin{eqnarray*}
 &&\mHsp(\im)^2\frac{e^2}{\epsi_0}\frac{m^{-\epsi/2}}{(4\pi)^{D/2}}\,(-1)^n\gn\int_0^1\dif x\int_0^1 \dif y\,\sqrt y\;
  \Bigg\{\Big[\pslp+\big(\bgp-\bgd x\big)y-\gn\big(p_0-d_0x\big)+m\Big]\nn
   &&\mHsp\times\Big[\wtp-\big(\bgp-\bgd
  x\big)y-\gn\big(p_0-d_0x\big)+m\Big]\,\Big(\frac{1}{w}\Big)
  +\halfS\big(3-\epsi-1/y)\;\frac{\Gamma(\epsi/2)}{w^{\epsi/2}}\Bigg\}
  \end{eqnarray*}
With Eqs (\ref{Dlim}) and (\ref{K}) we have in the limit
$\epsi\rarr0$
\begin{eqnarray}
  &-&K\gn\int_0^1\dif x\int_0^1 \dif y\,\sqrt y\;\Bigg\{\frac{NumC}{w}
  +\halfS\big(3-\epsi-1/y)\Big(\Delta-\ln(w/m^2)\Big)\Bigg\}\nn
  &=&-K\gn\int_0^1\dif x\int_0^1 \dif y\,\sqrt y\;\Bigg\{\frac{NumC}{w}
  +\Big(\halfS\big(3-1/y)\ln(w/m^2)-1\Big)\Bigg\}\Hsp
\end{eqnarray}
and after partial <integration
\begin{eqnarray}\label{C2}
  &-&K\gn\int_0^1\dif x\int_0^1 \dif y\,\sqrt y\;\Bigg\{\frac{NumC}{w}
  -(1-y)\,\frac{m^2}{w}\der{w}{y}-1\Bigg\}\Hsp
\end{eqnarray}
Here,
  \[NumC=\Big[\pslp+\big(\bgp-\bgd x\big)y-\gn\big(p_0-d_0x\big)+m\Big]
  \Big[\wtp-\big(\bgp-\bgd
  x\big)y-\gn\big(p_0-d_0x\big)+m\Big]\]
\begin{eqnarray*}
  &&w=(q^2y^2+(1-y)yq_0^2-sy=q_0^2y-\bq^2y^2-sy\nn
  &&\mhsp=y\big[(1-x)p_0+p_0'x\big]^2-y^2\big[(1-x)\bp+\bp'x\big]^2\nn
  &-&\big[(1-x)(p_0^2-\bp^2)+p_0\,'^2x-\bp\,'^2x-m^2\big]y
\end{eqnarray*}
and
 \begin{eqnarray}\label{dwdy}
  &&\der{w}{y}=\big[(1-x)p_0+p_0'x\big]^2-2y\big[(1-x)\bp+\bp'x\big]^2\nn
  &&\mhsp-\big[(1-x)(p_0^2-\bp^2)+p_0'^2x-\bp¨_0^2x-m^2\big]
  =\Delta_x-y\big[(1-x)\bp+\bp'x\big]^2\Hsp
\end{eqnarray}
with
\begin{eqnarray*}
  &&\mHsp\Delta_x=w/y=\big[(1-x)p_0+p_0'x\big]^2-y\big[(1-x)\bp+\bp'x\big]^2\nn
  &-&\big[(1-x)(p_0^2-\bp^2)+p_0'^2x-\bp'^2x-m^2\big]
\end{eqnarray*}
\begin{eqnarray}\label{Deltax}
  \boxed{\Delta_x
  =m^2-x(1-x)d_0^2+(1-x)xy\bd^2+(1-x)(1-y)\bp^2+x(1-y)\,\bp'^2}\hsp
\end{eqnarray}

The expression \eqqref{C2} then becomes
\begin{eqnarray}\label{C5}
  &-&K\gn\int_0^1\dif x\int_0^1 \dif y\,\sqrt y\;
  \frac{NumC-(1-y)\Big(\Delta_x-y\big[(1-x)\bp+x\bp'\big]^2\Big)-w}{w}\nn
  &=&-K\gn\int_0^1\dif x\int_0^1\frac{\dif y}{\sqrt y}\;
  \Bigg[ \frac{NumC-\Delta_x+y(1-y)\big[(1-x)\bp+x\bp'\big]^2}
  {\Delta_x}\Bigg]\Hsp
\end{eqnarray}

From \eqqref{C2}
\begin{eqnarray*}
  &&NumC=\Big[\gn p_0'-\bgp'+\big(\bgp-\bgd x\big)y-\gn\big(p_0-d_0x\big)+m\Big]\nn
  &\times&\Big[\bgp-\big(\bgp-\bgd x\big)y+\gn d_0x+m\Big]\nn
  &=&\Big[-\gn d_0(1-x)+m+\bgp(1-x)y-\bgp'(1-xy)\Big]\nn
  &\times&\Big[\gn d_0x+m+\bgp(1-y+xy)-\bgp'xy\Big]
 \end{eqnarray*}
\begin{eqnarray*}
  &&NumC=m^2-\gn md_0(1-2x)+m\bgd-d_0^2x(1-x)\nn
  &-&\gn d_0(1-x)\big[\bgp(1-y+xy)-\bgp'xy\big]
  +\big[\bgp(1-x)y-\bgp'(1-xy)\big]\gn d_0x\nn
  &-&\bgp\,\bgp'(1-x)xy^2-\bgp'\,\bgp(1-xy)(1-y+xy)\nn
  &+&(\bgp)^2(1-x)y(1-y+xy)+(\bgp')^2(1-xy)xy
\end{eqnarray*}
This can also be expressed
\begin{eqnarray*}
  &&NumC=m^2-\gn md_0(1-2x)+m\bgd-d_0^2x(1-x)\nn
  &-&\gn d_0\bgp(1-x)(1-y+2xy)+\gn d_0\bgp'x(1+y-2xy)\big]\nn
  &&\mHsp+(\bgp)^2(1-x)y(1-y)+(\bgp')^2(1-y)xy-\bgp'\,\bgp(1-y)
  -\bd^2 xy¨^2(1-x)
\end{eqnarray*}

With
\begin{eqnarray*}
  &&\big[(1-x)\bp+\bp'x\big]^2
  =-x(1-x)\bd^2+(1-x)\bp^2+x\bp'^2
\end{eqnarray*}
the numerator in \eqqref{C5} becomes
\begin{eqnarray*}
  &&-\gn R_x^0=NumC-\Delta_x+y(1-y)\big[(1-x)\bp+x\bp'\big]^2\nn
  &=&-\gn md_0(1-2x)+m\bgd\nn
  &-&\gn d_0\bgp\,(1-x)(1-y+2xy)+\gn d_0\bgp'x\,(1+y-2xy)\nn
  &-&\bgp'\,\bgp(1-y)-\bd^2 2xy(1-x)-\bp^2(1-x)(1-y)-\bp'^2x(1-y)
\end{eqnarray*}
The Coulomb contribution to the free-electron vertex function then
becomes
\begin{eqnarray}\label{VertCoul}
  \boxed{K\int_0^1\dif x\int_0^1\frac{\dif y}{\sqrt{y}}\,\frac{R_x^0}{\Delta
  x}}
\end{eqnarray}
which agrees with the result of Adkins~\cite{Adkins86} $R_x^0$
with
\[d=p-p'\rarr k;\;x\rarr u;\;y\rarr x\]

\subsection{Gaunt contribution}\label{sec:Gaunt}
For the \it{Gaunt part} we have
\begin{eqnarray}\label{G1e4}
  &&\mHsp-\frac{\im e^2}{\epsi_0}\intDD{k}\;
  \gii\frac{\psl\,'-\ksl+ m}{(p\,'-k)^2-m^2+\ime}\,
  \gn\frac{\psl-\ksl+m}{(p-k)^2-m^2+\ime}\,\gi\frac{1}{k^2+\ime}\Hsp
\end{eqnarray}
and, using the parametrization  \eqref{FI3} in Appendix \ref{sec:FI}
and the commutation rules in Appendix \ref{sec:gamma}, this becomes
\begin{eqnarray}\label{G2}
  &&\frac{\im e^2}{\epsi_0}\intDD{k}\;
  \gn\gii\frac{\wtp\,'-\wtk+ m}{(p\,'-k)^2-m^2+\ime}\,
  \frac{\psl-\ksl+m}{(p-k)^2-m^2+\ime}\,\gi\frac{1}{k^2+\ime}\nn
  &&\mHsp=\frac{2\im e^2}{\epsi_0}\intDD{k}\;\int_0^1\dif
  x\int_0^x\dif y\,\gn\frac{\gii(\wtp\,'-\wtk+ m)(\psl-\ksl+m)\gi}
  {\big[k^2+(p^2-2pk-m^2)x+(p'^2-p^2-2p'k+2pk)y\big]^3}\Hsp\nn
  &=&\frac{2\im e^2c}{\epsi_0}\intDim{k}\;\int_0^1\dif
  x\int_0^x\dif y\,\gn\frac{\gii(\wtp\,'-\wtk+ m)(\psl-\ksl+m)\,\gi}
  {\big[k^2+2kq+s\big]^3}
\end{eqnarray}
with $q=-(x-y)p-p'y$ and $s=(p^2-m^2)x-(p^2-p'^2)y$.

Applying the commutation rule $\nott A\gi=-\gi\!\!\nott A+2A^i$
and $\wt A\gi=-\gi\wt A-2A^i$ the numerator becomes
\begin{eqnarray}\label{Num}
  &&Num=\gii(\wtp\,'-\wtk+ m)(\psl-\ksl+m)\gi\nn
  &=&\Big[(-\wtp\,'+\wtk+m)\gii-2(p_i'-k_i)\Big]
  \Big[\gi(-\psl+\ksl+m)+2(p^i-k^i)\Big]\nn
    &=&\Big[(3-\epsi)(\wtp\,'-\wtk-m)-2(\bgp'-\bgk)\big](\psl-\ksl-m)\nn
  &+&2(\wt p\,'-\wt k-m)(\bgp-\bgk)+4(\pdpp-\pdk-\ppdk+\bk^2)\Hsp
\end{eqnarray}

We can now apply the formulas \eqref{DI4} to \eqref{DI6} in Appendix
\ref{sec:IntReg} with \makebox{$k\rarr-q=(x-y)p+p'y$} in
\eqqref{DI5} and first part of \eqqref{DI6} and $k^\mu k^\nu\rarr
-g^{\mu\nu}/2$ in the second part and
\begin{eqnarray}\label{w}
  &&w=q^2-s=[(x-y)p+p'y]^2-(p^2-m^2)x+(p^2-p'^2)y\hsp (y\rarr xu)\nn
  &&w/x=\Delta_y=x[(1-u)p+p'u]^2-(p^2-m^2)+(p^2-p'^2)u=w'x+w''\nn
  &&w'=[(1-u)p+p'u]^2\hsp w''=-(p^2-m^2)+(p^2-p'^2)u\nn
  &&\Delta_y=p^2(1-u)(x-xu-1)-p'^2x(1-xu)+(pp'+p'p)\,xu(1-u)+m^2\nn
  &=&m^2-ux(1-u)d^2-p^2(1-u)(1-x)-p'^2u(1-x)
\end{eqnarray}
This Agrees with Adkins' $\Delta_y$~\cite{Adkins86}.

The Gaunt part \eqref{G2} then becomes with the substitution
$y\rarr xu$

\begin{eqnarray}\label{Gaunt}
  \mhsp2K\gn\frac{1}{\Gamma(3)}\int_0^1x\,\dif x\int_0^1\,\dif u\,
  \Bigg\{\frac{\big[Num\big]_{k\rarr-q}}{w}
  +\Big[Num\Big]_{\big[k^\mu k^\nu\rarr-\halfS g^{\mu\nu}\big]}
  \frac{\Gamma(\epsi/2)}{w^{\epsi/2}}\Bigg\}\hsp
\end{eqnarray}
The evaluation will be made below together with the Gaunt-like
part of the scalar retardation.

\subsection{Scalar-retardation contribution}
The \bfit{scalar-retardation part} becomes similarly
\begin{eqnarray}\label{SR1}
  &&\mhsp-\frac{\im e^2}{\epsi_0}\intDD{k}\;
  \gi k_i \frac{\psl\,'-\ksl+ m}{(p\,'-k)^2-m^2+\ime}\,\gn\frac{\psl-\ksl+m}
  {(p-k)^2-m^2+\ime+\ime}\,\frac{\gamma^j k_j}{\bk^2}\frac{1}{k^2+\ime}\nn
  &&\mHsp=\frac{\im e^2}{\epsi_0}\intDD{k}\;
  \gn\gi k_i \frac{\wtp\,'-\wtk+m}{(p\,'-k)^2-m^2+\ime}\,\frac{\psl-\ksl+m}
  {(p-k)^2-m^2+\ime+\ime}\,\frac{\gamma^j k_j}{\bk^2}\frac{1}{k^2+\ime}\nn
  &&\mhsp=\frac{2\im e^2}{\epsi_0}\,\gn\intDim{k}\;\int_0^1\dif
  x\int_0^x\dif y\,\frac{\bgk(\wtp\,'-\wtk+m)(\psl-\ksl+m)\,\bgk}
  {\big[k^2+2kq+s\big]^3}\frac{1}{\bk^2}
\end{eqnarray}
with (as in the Gaunt case) $q=-(x-y)p-p'y$ and
$s=(p^2-m^2)x-(p^2-p'^2)y$. With the commutation rule
$\Asl\gi=-\gi\Asl+2A^i$ we have
  \[\Asl\,\gj k_j=-\gj k_j\Asl+2A^jk_j\,;\hhsp
  \Asl\,\bgk=-\bgk\,\Asl+2\bgk\,;\hhsp
  \wt A\,\bgk=-\bgk\,\wt A-2\bgk\]
  \[\bgk\Asl\,\bgk=-\bgk\,\bgk\Asl+2\bgk\,\bs{A}\bsdot\bk
  =\bk^2\Asl+2\bgk\,\bs{A}\bsdot\bk\]
  \[\bgk\ksl\,\bgk=\bk^2\ksl+2\bgk\,\bk^2=\bk^2\,\wtk\]
and the numerator in \eqqref{SR1} becomes
\begin{eqnarray}\label{GNum2}
  &&Num=-\bgk(\wtp\,'-\wtk+m)\big[\bgk(\psl-\ksl-m)
  -2(\pdk-\bk^2)\big]\nn
  &&=-\bk^2(\wtp\,'-\wtk-m)(\psl-\ksl-m)
  +2\bgk\,(\ppdk-\bk^2)(\psl-\ksl-m)\nn
  &+&2\bgk\,(\wtp'-\wtk+m)(\pdk-\bk^2)
\end{eqnarray}

\subsubsection{Gaunt-like part}
The parts involving $\bk^2$ can be evaluated as the Gaunt part,
\begin{eqnarray*}
  \mHsp&&-(\wtp\,'-\wtk-m)(\psl-\ksl-m)-2\bgk\,(\psl-\ksl-m)
  -2\bgk\,(\wtp\,'-\wtk+m)-2\ppdk+2\pdk
\end{eqnarray*}
which together with the Gaunt part \eqqref{Num}
\begin{eqnarray}\label{Num2}
 &&\Big[(3-\epsi)(\wtp\,'-\wtk-m)-2(\bgp'-\bgk)\Big](\psl-\ksl-m)\nn
  &+&2(\wt p\,'-\wt k-m)(\bgp-\bgk)+4(\pdpp-\pdk-\ppdk+\bk^2)\Hsp
\end{eqnarray}
gives (for the time being omitting $\epsi$)
\begin{eqnarray}\label{NumG}
  &&\mhsp NumG=2\Big[(\gn p_0'-\wtk-m)(\psl-\ksl-m)
  +(\wt p\,'-\wt k-m)\bgp\nn
  &&\mHsp-\bgk\,(\wtp\,'-\wtk+m)-(\wt p\,'-\wt k-m)\bgk -\ppdk+\pdk
  +2(\pdpp-\pdk-\ppdk+\bk^2)\Big]\nn
  &=&2\Big[(\gn p_0'-\wt k-m)(\psl-\ksl-m)
  +(\wt p\,'-\wt k-m)\bgp\nn
  &-&\bgk\,\bgp'-\bgp'\,\bgk+2(\bgk)^2\nn
  &-&\ppdk+\pdk+2(\pdpp-\pdk-\ppdk+\bk^2)\Big]\nn
  &=&2\Big[(\gn p_0'-\gn k_0-\bgk-m)(\gn p_0-\bgp-\gn
  k_0+\bgk-m)\nn
  &+&(\gn p_0'+\bgp'-\gn k_0-\bgk-m)\bgp+2\pdpp-\pdk-\ppdk)\Big]\nn
  &=&2\Big[(p_0'-k_0)(p_0-k_0)-m(\gn p_0+\gn p_0'-2\gn k_0-m)
  +\bgp\,'\bgp\nn
  &-&\bgk(\gn p_0+\gn p_0'-2\gn k_0)+2\pdpp-\pdk-\ppdk+\bk^2)\Big]
\end{eqnarray}

We can now evaluate the entire Gaunt-like contribution by
replacing $Num$ in \eqqref{Gaunt} by $NumG$,
\begin{eqnarray}\label{Gaunt-like}
  \mhsp K\gn\int_0^1x\,\dif x\int_0^1\,\dif u\,
  \Bigg\{\frac{\big[NumG\big]_{k\rarr-q}}{w}
  +\Big[NumG\Big]_{\big[k^\mu k^\nu\rarr-\halfS g^{\mu\nu}\big]}
  \frac{\Gamma(\epsi/2)}{w^{\epsi/2}}\Bigg\}\hsp
\end{eqnarray}
and adding the contribution due to $\epsi$ in \eqqref{Num2}. The
substitution in the second part gives
   \[k_0^2\rarr-\halfS;\hsp \bk^2=-\gi k_i\,\gj\,k_j
   \rarr\halfS\gi\gii=\halfS(3-\epsi)\]
and
  \[\Big[NumG\Big]_{\big[k^\mu k^\nu\rarr-\halfS g^{\mu\nu}\big]}
  =2-\epsi\]
  \[\frac{\Gamma(\epsi/2)}{(w/m^2)^{\epsi/2}}
  =\Delta-\ln \Big(\frac{w}{m^2}+\cdots\Big)\]
The omitted $\epsi$-dependent contribution in \eqqref{NumG} is
$-\epsi$. This gives the Gaunt-like contribution
\eqref{Gaunt-like}
\begin{eqnarray}\label{GSR}
  K\gn\int_0^1\dif u\int_0^1x\,\dif x\Bigg[\frac{\big[NumG\big]_{k\rarr xq_A}}{w}
  +2(1-\epsi)\Big[\Delta-\ln
  \Big(\frac{w}{m^2}\Big)\Big]\Bigg]\hsp
\end{eqnarray}
Partial integration of the log term yields (where according to Eq.
\ref{w} $w=w'x^2+w''x$ and $\der{w}{x}=2xw'+w''$)
\begin{eqnarray}
  &&\int_0^1x\,\dif x\ln\Big(\frac{w}{m^2}\Big)=
  \Big[\frac{x^2}{2}\ln\Big(\frac{w}{m^2}\Big)\Big]_0^1
  -\int_0^1\dif x\,\frac{x^2}{2}\,\frac{m^2}{w}\,\der{w}{x}\nn
  &=&\half \ln\Big(\frac{w'+w''}{m^2}\Big)
  -\int_0^1\dif x\,\frac{x^2}{2}\,\frac{m^2}{w}\,(2xw'+w'')
\end{eqnarray}
The result \eqqref{GSR} then becomes
\begin{eqnarray}\label{GSR2}
  &&\mHsp K\gn\int_0^1\dif u\Bigg[\int_0^1x\,\dif x\,
  \frac{\big[NumG\big]_{k\rarr xq_A}+x(2xw'+w'')}{w}+\Delta-2
  -\ln\Big(\frac{w'+w''}{m^2}\Big)\Bigg]\Hsp
\end{eqnarray}
The constant factor can be moved to the numerator as
$-4w=-4x^2w'-4xw''$, which yields with $w=x\Delta_y$
\begin{eqnarray}\label{GSR3}
  &&\mhsp K\gn\int_0^1\dif u\Bigg[\int_0^1\dif x\,
  \frac{\big[NumG\big]_{k\rarr xq_A}-2x^2w'-3xw''}{\Delta_y}+\Delta
  -\ln\Big(\frac{w'+w''}{m^2}\Big)\Bigg]\Hsp
\end{eqnarray}
Here, the numerator becomes
\begin{eqnarray}\label{Ry3}
  &&2R_y^0=\big[NumG\big]_{k\rarr xq_A}-x(2xw'+3w'')\nn
  &&\mHsp=2\Big[(p_0'-k_0)(p_0-k_0)-m(\gn p_0+\gn p_0'-2\gn k_0-m)
  +\bgp\,'\bgp\nn
  &&\mHsp-\bgk(\gn p_0+\gn p_0'-2\gn
  k_0)+2\pdpp-\pdk-\ppdk+\bk^2)\Big]_{k\rarr xq_A}\nn
  &-&x(2xw'+3w'')
\end{eqnarray}
or with
\begin{eqnarray}\label{ww'}
  w'+w''=[(1-u)p+p'u]^2-(p^2-m^2)+(p^2-p'^2)u=m^2-u(1-u)(p-p')^2\hhsp
\end{eqnarray}
\begin{eqnarray}\label{Ry4}
  &&R_y^0=(p_0'-xq_A^0)(p_0-xq_A^0)-m(\gn p_0+\gn p_0'-2\gn xq_A^0-m)\nn
  &+&\bgp\,'\bgp-x\bgq_A(\gn p_0+\gn p_0'-2\gn xq_A^0)\nn
  &+&2\pdpp-x\pdq_A-x\ppdq_A +x^2\bq_A^2\nn
  &-&x^2\big[(q_A^0)^2-\bq_A^2\big]+\frac{3x}{2}\big[p^2(1-u)+p'^2u-m^2\big]
\end{eqnarray}
Here, $q_A=(1-u)p+up'$ and
\begin{eqnarray*}
  &&\mhsp x\pdq_A+x\ppdq_A=x(\bp+\bp')\bsdot\big[(1-u)\bp+u\bp'\big]
  =x(1-u)\bp^2+xu\bp'^2+x\pdpp
\end{eqnarray*}
The scalar part of the last line in \eqqref{Ry4} becomes
\begin{eqnarray*}
  &&-x^2\big[(1-u)p_0+up_0'\big]^2+\frac{3x}{2}\big[(1-u)p_0^2
  +up_0'^2-m^2\big]\nn
  &=&-\frac{3x}{2}m^2+x^2u(1-u)d_0^2+\big[-x^2(1-u)^2+\frac{3x}{2}(1-u)-x^2u(1-u)\big]p_0^2\nn
  &+&\big[-x^2u^2+\frac{3x}{2}u-x^2u(1-u)\big]p_0'^2\nn
  &=&-\frac{3x}{2}\,m^2+x^2u(1-u)d_0^2+\halfS
  x(1-u)(3-2x)p_0^2+\halfS xu(3-2x)p_0'^2
\end{eqnarray*}
and the double vector part of the last two lines
\begin{eqnarray*}
  &&\mHsp-\big[1-x/2+2x^2u(1-u)\big]\bd^2+\big[2x^2(1-u)^2-\frac{5x}{2}(1-u)+1-x/2+2x^2u(1-u)\big]\bp^2\nn
  &+&\big[2x^2u^2-\frac{5x}{2}u+1-x/2+2x^2u(1-u)\big]\bp'^2\nn
  &=&-\big[1-x/2+2x^2u(1-u)\big]\bd^2+\big[1-x/2-\frac{x}{2}(1-u)(5-4x)\big]\bp^2\nn
  &+&\big[1-x/2-\halfS xu(5-4x)\big]\bp'^2
\end{eqnarray*}
This gives (times $\gn$)
\begin{eqnarray}\label{RyFinal}
  R_y^0&=&(1-\frac{3x}{2}m^2)+(p_0'-xq_A^0)(p_0-xq_A^0)-m(\gn p_0+\gn p_0'-2\gn xq_A^0)\nn
  &&+x^2u(1-u)d_0^2+\halfS
  x(1-u)(3-2x)p_0^2+\halfS xu(3-2x)p_0'^2\nn
  &+&\bgp\,'\bgp-x\bgq_A(\gn p_0+\gn p_0'-2\gn xq_A^0)\nn
  &-&\big[1-x/2+2x^2u(1-u)\big]\bd^2+\big[1-x/2-\frac{x}{2}(1-u)(5-4x)\big]\bp^2\nn
  &+&\big[1-x/2-\halfS xu(5-4x)\big]\bp'^2
\end{eqnarray}
The Gaunt-like contribution then becomes
\begin{eqnarray}\label{VertGaunt}
  \boxed{K\gn\Bigg[\Delta+\int_0^1\dif x\int_0^1\dif y\,\frac{2R_y^0}{\Delta
  y}-\int_0^1\dif u\,\ln\frac{m^2-w-w'}{m^2}\Bigg]}
\end{eqnarray}
where $w+w'$ is given by \eqqref{ww'}. This agrees with the result
of Adkins~\cite{Adkins86}.

\subsubsection{Non-Gaunt-like part}
For the remaining "non-Gaunt-like" part of the scalar retardation
\eqref{SR1} we apply the formulas \eqref{DI6k} and \eqref{DI7k} in
Appendix \ref{sec:IntReg}, leading to an expression of the type
after the substitution $y\rarr xu$
\begin{eqnarray}\label{SRNG}
  &&2K\gn\int_0^1x\,\dif x\int_0^1\dif u\int_0^1\dif
  z\,z^{3/2}\,\frac{1}{\Gamma(3)}\Bigg[\Gamma(2)\frac{N_Z'}{w^2}
  +\frac{N_Z}{w}\Bigg]\nn
  &=&K\gn\int_0^1x\,\dif x\int_0^1\dif u\int_0^1\dif
  z\,z^{3/2}\Bigg[\frac{N_Z'}{w^2}
  +\frac{N_Z}{w}\Bigg]
\end{eqnarray}
Here, with $y=xu$ and $q=x[(1-u)p+up']$
 \begin{eqnarray*}
  &w&=q^2z^2+(1-z)z\,q_0^2-sz=-z^2\bq^2+\,q_0^2-sz\nn
  &s&=(p^2-m^2)x-(p^2-p'^2)y\
\end{eqnarray*}
\begin{eqnarray}\label{wz}
  &w&=-x^2z^2\Big[(1-u)\bp+u\bp'\big]^2+zx^2\Big[(1-u)p_0+up_0'\big]^2\nn
  &-&xz(p^2-m^2)+zxu(p^2-p'^2)=xz\Delta_z
\end{eqnarray}
This gives
\begin{eqnarray}\label{Deltaz}
  &\Delta_z&=-xz\big[(1-u)\bp+u\bp'\big]^2+x\big[(1-u)p_0+up_0'\big]^2\nn
  &-&(p^2-m^2)+u(p^2-p'^2)\nn
  &=&m^2+xuz(1-u)\bd^2-xu(1-u)d_0^2+(1-u)(1-xz)\bp^2+u(1-xz)\bp'^2\nn
  &-&(1-x)(1-u)p_0^2-u(1-x)p_{0}^{'2}
\end{eqnarray}
which agrees with Adkins with $x\rarr s\,;\;u\rarr u\,;\;z\rarr
x$~\cite{Adkins86}.

We can now express the non-Gaunt part \eqref{SRNG}  as
\begin{eqnarray}\label{SRNG1}
  K\gn\int_0^1\,\dif x\int_0^1\dif u\int_0^1\dif
  z\,z^{1/2}\Bigg[\frac{N_Z'}{xz(\Delta_z)^2}
  +\frac{N_Z}{\Delta_z}\Bigg]
\end{eqnarray}

The non-Gaunt-like part of the numerator \eqqref{GNum2} is
\begin{eqnarray*}
 NumNG= 2\Big[\bgk\,\bp'\bsdot\bk\,(\psl-\gn k_0-m)
  +\bgk\,\bp\bsdot\bk\,(\wtp\,'-\gn k_0+m)\Big]
\end{eqnarray*}
In the first part of \eqqref{DI6k} and \eqqref{DI7k} we then make
the replacements (c.f. Coulomb contribution)
  \[k^i\rarr-q^iz=q_A^ixz\,;\hsp\bk\rarr\bq_Axz\,;\hsp k_0\rarr-q_0=q_A^0x\]
  \[\bgk\gn k_0\,\bp\bsdot\bk
  \rarr  \bgq_A\,\gn q_A^0\,\bp\bsdot\bq_A\;x^3z^2\]
Then
\begin{eqnarray}
  N_z'=\Big[NumNG\Big]_{k_0\rarr xq_a^0;\,\bk\rarr xz\bq_A}
  =2x^2z^2R_z'^0
\end{eqnarray}
which gives
\begin{eqnarray*}
  &&R_z'^0=\bgq_A\,\bp'\bsdot\bq_A\,(\psl-m)+\bgq_A\,\bp\bsdot\bq_A\,(\wtp\,'+m)\nn
  &-&x\,\bgq_A\,\gn q_A^0\big(\bp'\bsdot\bq_A+\bp\bsdot\bq_A\big)
\end{eqnarray*}
or
\begin{eqnarray}\label{Rz'}
  &R_z'^0=&m\,\bgq_A\,(\bp\bsdot\bq_A-\bp'\bsdot\bq_A)\rarr m\,
  \bgq_A\,\bd\bsdot\bq_A\nn
  &+&\bgq_A\,(\bp'\bsdot\bq_A\,\gn p_0+\bp\bsdot\bq_A\,\gn p_0')\nn
  &+&\bgq_A\,(\bp\bsdot\bq_A\,\bgp'-\bp'\bsdot\bq_A\,\bgp)\nn
  &-&x\,\bgq_A\,\gn q_A^0\big(\bp'\bsdot\bq_A+\bp\bsdot\bq_A\big)
\end{eqnarray}
The third line can be expressed
\begin{eqnarray*}
  &&\bgq_A\,(\bp\bsdot\bq_A\,\bgp'-\bp'\bsdot\bq_A\,\bgp)\nn
  &=&\big[(1-u)\bgp+u\bgp'\big]\Big(\bgp'\big[(1-u)\bp^2+u\pdpp\big]
  -\bgp\big[(1-u)\pdpp+u\bp'^2\big]\Big)\nn
  &=&C_1+C_2+C_3
\end{eqnarray*}
where
\begin{eqnarray*}
  \left\{\begin{array}{l}
  C_1=\bgp\,\bgp'\big[(1-u)^2\bp^2+u(1-u)\pdpp\big]=\bgp\,\bgp'\,A\\
  C_2=-\bgp'\,\bgp\big[u(1-u)\pdpp+u^2\bp'^2\big]=-\bgp'\,\bgp\,B\nn
  C_3=\pdpp\big[(1-u)^2\bp^2-u^2\pdpp\big]=\pdpp(A-B)
  \end{array}\right.
\end{eqnarray*}
with $A+B=\bq_A^2$. $C_1$ can be reexpressed as
  \[C_1=-(\bgp'\,\bgp+2\pdpp)\,A\]
which leads to
  \[C_1+C_2+C_3=-\bgp'\,\bgp\,(A+B)-\,\pdpp(A+B)\]
or
\begin{eqnarray}
  \bgq_A\,(\bp\bsdot\bq_A\,\bgp'-\bp'\bsdot\bq_A\,\bgp)
  =-(\bgp'\,\bgp+\pdpp)\;\bq_A^2\hsp
\end{eqnarray}

In the second part of \eqqref{DI6k} we make the replacement
\[k^ik^j\rarr-\halfS g^{ij}\,;\hsp\bgk\,\bp\bsdot\bk\rarr\halfS\bgp\] and in
\eqqref{DI7k}
  \[ k^ik^0k^j\rarr-\halfS g^{ij}q_0\]
  or
\[\bgk\gn k_0\,\bp\bsdot\bk=\gi k_i\,\gn k_0\,p^jk_j\,\rarr
-\halfS \gj p_j\,\gn q_0=+\halfS\bgp\,\gn q_A^0x \] The numerator
in the second part of \eqqref{SRNG}is then
\begin{eqnarray}
  N_z=\Big[NumNG\Big]_{k^ik^j\rarr-\halfS g^{ij};\,k^ik^0k^j\rarr-\halfS g^{ij}q_0}
  =R_z^0
\end{eqnarray}
which gives
\begin{eqnarray}\label{NumSR2}
  R_z&=&\bgp'(\psl-m)+\bgp(\wtp\,'+m)-\bgp\,\gn q_A^0x-\bgp'\,\gn q_A^0x\nn
  &=&m\bgd+(\bgp'\,p_0+\bgp\,\p_0\,')+(\bgp+\bgp')\gn q_A^0x\nn
  &+&\bgp\,\bgp'-\bgp'\,\bgp\nn
  &=&m\bgd+(\bgp'\,p_0+\bgp\,\p_0\,')+(\bgp+\bgp')\gn q_A^0x\nn
  &+&2\bp\bsdot\bp'+2\bgp\,\bgp'
 \end{eqnarray}
The non-Gaunt-like contribution to the free-electron vertex
function then becomes
\begin{eqnarray}\label{VertNonGaunt}
  \boxed{K\gn\int_0^1\dif x\int_0^1\dif u\int_0^1\dif z\,\sqrt{z}\,
  \Bigg[\frac{2xzR_z'^0}{(\Delta z)^2}+\frac{R_z^0}{\Delta z}\Bigg]}
\end{eqnarray}
which agrees with Adkins~\cite{Adkins86} with
\[x\rarr s;\, z\rarr x\]

We have now verified the formulas of Adkins for dimensional
regularization in the Coulomb gauge of the free-electron self
energy and vertex function for $\mu=0$.

\section*{Acknowledgements}
The author is much obligated to his collaborators, Sten
Salomonson, Daniel Hedendahl, and Johan Holmberg, for many
valuable discussions.
\appendix

\section{Relations for the alpha and gamma matrices}\label{sec:gamma}
Taken from ref.~\cite[App. D3]{ILBook11}.\footnote{Note misprint
in the
first formula of Eq. (D.58).}\\
The gamma matrices satisfy the following anti-commutation rule:
\begin{eqnarray}  \label{gammaAC}
   \gnu\gmu+\gmu\gnu&=&2g^{\mu\nu}\nn
   \nott A\nott B+\nott B\nott A&=&2AB
\end{eqnarray}
where $\nott{A}$ is defined
\begin{equation}\label{Aslash}
 \!\!\nott{A}=\gamma^\mu \hat{A}_\mu
\end{equation}
This leads to
\begin{eqnarray}  \label{gammaRel}
   \gamma^\mu\gamma^\nu\gamma_\mu&=&-2\gamma^\nu\nn
   \gamma^\mu\nott A\gamma_\mu&=&-2\nott A\nn
   \gamma^\mu\gamma_\mu&=&4\nn
   \gamma^\mu\gamma^\nu\gamma_\mu&=&-2\gamma_\mu\nn
   \gamma^\mu\nott A\gamma_\mu&=&-2\nott A\nn
   \gn\gamma_0&=&\gn\gn=1\nn
   \gs\gn&=&\gn\wts\nn
   \nott A\gn&=&\gn\wt{A}\nn
   \gn\gs\gn&=&\wts\nn
   \gn\nott A\,\gn&=&\wt{A}\nn
   \gn\gs\gt\gn&=&\wts\wtg^\tau\nn
   \gn\nott A\nott\,B\,\gn&=&\wt{A}\wt{B}\nn
   \gn\gb\gs\gt\gn&=&\wtg^\beta\wts\wtg^\tau\nn
   \gn\nott A\nott\,B\,\nott \,C\,\gn&=&\wt{A}\wt{B}\wt{C}
\end{eqnarray}
where $\wt{A}=\gn A_0-\gi A_i=\gn+\bgd\bs{A}$ and
  \[\wts=\left\{\begin{array}{c}\gs (\sigma=1)\\-\gs (\sigma=1,2,3
  \end{array}\right.\]

With the number of dimensions being equal to $4-\epsi$, to be used
in dimensional regularization, the relations become
\begin{eqnarray}  \label{gammaDim}
   \gmu\gamma_\mu&=&4-\epsi\nn
   \gmu\gs\gamma_\mu&=&-(2-\epsi)\gs\nn
   \gmu\nott A\gamma_\mu&=&-(2-\epsi)\nott A\nn
   \gmu\gs\gt\gamma_\mu&=&4g^{\sigma\tau}-\epsi\gs\gt\nn
   \gmu\nott A\,\nott B\gamma_\mu&=&4AB-\epsi\nott A\nott B\nn
   \gmu\gb\gs\gt\gamma_\mu&=&-2\gt\gs\gb+\epsi\gb\gs\gt\nn
   \gii\gi&=&3-\epsi\nn
   \gii\gs\gi&=&-(2-\epsi)\gs-\wts\nn
   \gii\nott A\gi&=&-(2-\epsi)\nott A-\wt A\nn
   \gii\gs\gt\gi&=&4g^{\sigma\tau}-\wts\wtg^\tau-\epsi\gs\gt\nn
   \gii\nott A\,\nott B\,\gi&=&4AB-\wt{A}\wt{B}-\epsi\nott A\nott B\nn
   \gii\gb\gs\gt\gi&=&-2\gt\gs\gb-\wtg^\beta\wts\wtg^\tau+\epsi\gb\gs\gt\nn
   \gii\nott A\nott\,B\,\nott \,C\,\gi&=&-2\nott C\,\nott B\,\nott A\,-\wt{A}\wt{B}\wt{C}
   +\epsi\nott A\,\nott B\,\nott C
\end{eqnarray}

\section{Formulas for dimensional regularization}\label{sec:IntReg}
Taken from ref.~\cite[App. G2]{ILBook11}.

Following the book by Peskin and Schroeder~\cite{PS95}, we can by
means of \it{Wick rotation} evaluate the integral
\begin{eqnarray*}
   &&\intDim{l}\,\frac{1}{(l^2-\Delta)^m}=\im(-1)^m\intDim{l}\,
   \frac{1}{(l_E^2+\Delta)^m}\nn
   &=&\im(-1)^m\int\frac{\dif\Om_D}{(2\pi)^4}\int_0^\infty\dif l_E^0\,
   \frac{l_E^{D-1}}{(l_E^2+\Delta)^m}
\end{eqnarray*}
We have here made the replacements $l^0=\im l_E^0$ and
$\bs{l}=\bs{l_E}$ and rotated the integration contour of $l_E$
$90^o$, which with the positions of the poles should give the same
result. The integration over $\dif^D l_E$ is separated into an
integration over the D-dimensional sphere $\Om_D$ and the linear
integration over the component $l_E^0$. This corresponds in three
dimensions to the integration over the two-dimensional angular
coordinates and the radial coordinate (see below).

\begin{equation}\label{DI1}
   \intDim k\,\frac{1}{(k^2+s+\ime)^n}
   =\frac{\im(-1)^n}{4\pi^{D/2}}\frac{\Gamma(n-D/2)}{\Gamma(n)}\frac{1}{s^{n-D/2}}
\end{equation}
\begin{equation}\label{DI2}
   \int\dif^4 k\,\frac{k^\mu}{(k^2+s+\ime)^n}=0
\end{equation}
\begin{equation}\label{DI3}
   \intDim k\,\frac{k^\mu k^\nu}{(k^2+s+\ime)^n}
   =\frac{\im(-1)^n}{4\pi^{D/2}}\frac{\Gamma(n-D/2-1)}{\Gamma(n)}\frac{1}{s^{n-D/2-1}}
\end{equation}

\subsubsection*{Covariant gauge}
Compared to Adkins~\cite{Adkins83} Eqs (A1a), (A3), (A5a):
\begin{eqnarray}&&p\rarr-q;\;M^2\rarr-s;\;\om\rarr D/2;\;\alpha\rarr n;\;\xi\rarr n;
  \;Q=p\rarr-q;\;A_{\mu\nu}\rarr g_{\mu\nu};\;\nn
  && \Delta\rarr w=q^2-s
\end{eqnarray}
\begin{equation}\label{DI4}
   \intDim k\,\frac{1}{(k^2+2kq+s+\ime)^n}
   =\frac{\im(-1)^n}{(4\pi)^{D/2}}\frac{1}{\Gamma(n)}\frac{\Gamma(n-D/2)}{w^{n-D/2}}
\end{equation}
\begin{eqnarray}\label{DI5}
   \mhsp\intDim k\,\frac{k^\mu}{(k^2+2kq+s+\ime)^n}
   =-\frac{\im(-1)^n}{(4\pi)^{D/2}}\frac{1}{\Gamma(n)}\,q^\mu\,
   \frac{\Gamma(n-D/2)}{{w^{n-D/2}}}\Hsp
\end{eqnarray}
\begin{eqnarray}\label{DI6}
   &&\mHsp\intDim k\,\frac{k^\mu k^\nu}{(k^2+2kq+s+\ime)^n}
   =\frac{\im(-1)^n}{(4\pi)^{D/2}}\,\frac{1}{\Gamma(n)}
   \Bigg[q^\mu q^\nu\,\frac{\Gamma(n-D/2)}{w^{n-D/2}}\nn
   &&-\frac{g^{\mu\nu}}{2}\,\frac{\Gamma(n-1-D/2)}{w^{n-1-D/2}}\Bigg]
\end{eqnarray}

\subsubsection*{Non-covariant gauge}
Compared to Adkins~\cite{Adkins83} Eqs (A1b), (A4), (A5b):
\begin{eqnarray*}
  &&p\rarr-q;\;M^2\rarr-s;\;\om\rarr D/2;\;\alpha\rarr n;\; \beta\rarr1;\;\xi\rarr
  n+1;\;\bk^2\rarr-\bk^2;\;\nn
  &&Q=py\rarr-qy;\;
  A_{\mu\nu}\rarr g_{\mu\nu}+\delta_{\mu,0}\delta_{\nu,0}\frac{1-y}{y};\;
  (AQ)_\mu\rarr-q^\mu y-\delta_{\mu0}\,(1-y)q_0\nn
  &&\Delta\rarr w= q^2y^2+(1-y)yq_0^2-sy+\lambda^2(1-y)=
  -\bq^2y^2+yq_0^2-sy+\lambda^2(1-y)
\end{eqnarray*}
\begin{eqnarray}\label{DI4k}
   \mHsp&&\intDim k\,\frac{1}{(k^2+2kq+s+\ime)^n}\frac{1}{\bk^2-\lambda^2}\nn
   &&=\frac{\im(-1)^n}{(4\pi)^{D/2}}\,\frac{1}{\Gamma(n+1)}
   \int_0^1\dif y\,y^{n-1-1/2} \,\frac{\Gamma(n+1-D/2)}{w^{n+1-D/2}}
\end{eqnarray}
\begin{eqnarray}\label{DI5k}
   &&\mhsp\intDim k\,\frac{k^\mu}{(k^2+2kq+s+\ime)^n}\frac{1}{\bk^2-\lambda^2}
   =-\frac{\im(-1)^n}{(4\pi)^{D/2}}\,\frac{1}{\Gamma(n)}\nn
   &\times&\int_0^1\dif y\,y^{n-1-1/2}\big[q^\mu y+\delta_{\mu,0}\,q_0(1-y)\big]
   \,\frac{\Gamma(n+1-D/2)}{w^{n+1-D/2}}\hsp
\end{eqnarray}
  \[\ksl\rarr-\qsl\,y-\gn q_0(1-y)=\bgq\,y-\gn q_0\]
\begin{eqnarray}\label{DI6k}
   &&\mhsp\intDim k\,\frac{k^\mu k^\nu}{(k^2+2kq+s+\ime)^n}\frac{1}{\bk^2-\lambda^2}
   =\frac{\im(-1)^n}{(4\pi)^{D/2}}\,\frac{1}{\Gamma(n)}\nn
   &&\mhsp\times\int_0^1\dif y\,y^{n-1-1/2}
   \Bigg[\Big\{\big[q^\mu y+\delta_{\mu,0}\,q_0(1-y)\big]
   \big[q^\nu y+\delta_{\nu,0}\,q_0(1-y)\big]\Big\}
   \frac{\Gamma(n+1-D/2)}{w^{n+1-D/2}}\nn
   &&-\half\Big\{\big[g^{\mu\nu} +\delta_{\mu,0}\delta_{\nu,0}(1-y)/y\big]\Big\}
   \frac{\Gamma(n-D/2)}{w^{n-D/2}}\Bigg]
\end{eqnarray}
$\ksl\rarr\bgq\,y-\gn q_0$ in first part and
   $\ksl\ksl\rarr-\half\big[\gmu \gamma_\mu+(1-y)/y\big]$
   in second.
\begin{eqnarray}\label{DI7k}
   &&\intDim k\,\frac{k^i k^\mu k^j}{(k^2+2kq+s+\ime)^n}\frac{1}{\bk^2-\lambda^2}
   =-\frac{\im(-1)^n}{(4\pi)^{D/2}}\,\frac{1}{\Gamma(n)}\nn
   &\times&\int_0^1\dif y\,y^{n-1-1/2}
   \Bigg[\Big\{q^i q^\mu q^j y^3+q^i q^\mu q^j\delta_{\mu0} (1-y) y^2\Big\}
   \frac{\Gamma(n+1-D/2)}{w^{n+1-D/2}}\nn
   &&\mhsp+\half\Big\{y\big(g^{i\mu}q^j+g^{\mu
   j}q^i+g^{ji}q^\mu\big)-
   \delta_{\mu0}\,g^{ij}\,q_0(1-y)\Big\}\frac{\Gamma(n-D/2)}{w^{n-D/2}}\Bigg]\hsp
\end{eqnarray}

\section{Feynman integrals}\label{sec:FI}
Taken from ref.~\cite[App. J1]{ILBook11}\\
In this Appendix we shall give some integrals, which simplify many
QED calculations considerably (see the books of Mandl and
Shaw~\cite[Ch. 10]{MS84} and Sakurai~\cite[App. E]{Sak67}, and we
shall start by deriving some formulas due to Feynman.

We start with the identity
\begin{equation}  \label{FI1}
   \frac{1}{ab}=\frac{1}{b-a}\int_a^b \frac{\dif t}{t^2}
\end{equation}
With the substitution $t=b+(a-b)x$ this becomes
\begin{equation}  \label{FI2}
   \frac{1}{ab}=\int_0^1 \frac{\dif x}{[b+(a-b)x]^2}
   =\int_0^1 \frac{\dif x}{[a+(b-a)x]^2}
\end{equation}
Differentiation with respect to $a$, yields
\begin{equation}  \label{FI2A}
   \frac{1}{a^2b}=2\int_0^1 \frac{x\dif x}{[b+(a-b)x]^3}
\end{equation}
Similarly, we have
\begin{eqnarray}  \label{FI3}
   \frac{1}{abc}&=&2\int_0^1 \dif x\int_0^x \dif
   y\,\frac{1}{[a+(b-a)x+(c-b)y]^3}\nn
   &=&2\int_0^1 \dif x\int_0^{1-x} \dif y\,\frac{1}{[a+(b-a)x+(c-a)y]^3}
\end{eqnarray}

Next we consider the integral
\begin{equation*}
   \int\dif^4 k\,\frac{1}{(k^2+s+\ime)^3}
   =4\pi\int \ka^2\dif\ka\int_{-\infty}^\infty\frac{\dif k_0}{(k^2+s+\ime)^3}
\end{equation*}
The second integral can be evaluated by starting with
\begin{equation*}
   \int_{-\infty}^\infty\frac{\dif k_0}{k_0^2-\ka^2+s+\ime}
   =\frac{\im\pi}{\sqrt{\ka^2-s}}
\end{equation*}
evaluated by residue calculus, and differentiating twice with
respect to $s$. The integral then becomes
\begin{equation}\label{FI4}
   \int\dif^4 k\,\frac{1}{(k^2+s+\ime)^3}
   =\frac{3\im\pi^2}{2}\int \frac{\ka^2\dif\ka}{(\ka^2+s)^{5/2}}=\frac{\im\pi^2}{2s}
\end{equation}
The second integral can be evaluated from the identity
\begin{equation*}
   \frac{x^2}{(x^2+s)^{5/2}}=\frac{1}{(x^2+s)^{3/2}}
   -\frac{s}{(x^2+s)^{5/2}}
\end{equation*}
and differentiating the integral
\begin{equation*}
   \int\frac{\dif x}{\sqrt{x^2+s}}=\ln\big(x+\sqrt{x^2+s}\big)
\end{equation*}
yielding
\begin{equation*}
   \int\frac{x^2}{(x^2+s)^{5/2}}=\frac{1}{3s}
\end{equation*}

For symmetry reason we find
\begin{equation}\label{FI5}
   \int\dif^4 k\,\frac{k^\mu}{(k^2+s+\ime)^3}=0
\end{equation}
Differentiating this relation with respect to $k_\nu$, leads to
\begin{equation}\label{FI6}
   \int\dif^4 k\,\frac{k^\mu k^\nu}{(k^2+s+\ime)^4}=
   \frac{g^{\mu\nu}}{3}\int\dif^4 k\,\frac{1}{(k^2+s+\ime)^3}=\frac{\im\pi^2g^{\mu\nu}}{6s}
\end{equation}
using the relation ~\cite[Eq. (A4)]{ILBook11}.

By making the replacements
  \[k\Rarr k+q  \quad \rm{qnd} \quad s\Rarr s-q^2\]
the integrals \eqref{FI4} and \eqref{FI5} lead to
\begin{equation}\label{FI7}
   \int\dif^4 k\,\frac{1}{(k^2+2kq+s+\ime)^3}=\frac{\im\pi^2}{2(s-q^2 )}
\end{equation}
\begin{equation}\label{FI8}
   \mhsp\int\dif^4 k\,\frac{k^\mu}{(k^2+2kq+s+\ime)^3}
   =-\int\dif^4 k\,\frac{q^\mu}{(k^2+2kq+s+\ime)^3}
   =-\frac{\im\pi^2 q^\mu}{2(s-q^2 )}
\end{equation}
Differentiating the last relation with respect to $q_\nu$, leads
to
\begin{equation}\label{FI9}
   \int\dif^4 k\,\frac{k^\mu k^\nu}{(k^2+2kq+s+\ime)^4}
   =\frac{\im\pi^2}{12}\Big[\frac{g^{\mu\nu}}{s-q^2}+
   \frac{2q^\mu q^\nu}{(s-q^2)^2}\Big]\quad ??
\end{equation}

Differentiating the relation \eqref{FI7} with respect to $s$,
yields
\begin{equation}\label{FI10}
   \int\dif^4 k\,\frac{1}{(k^2+2kq+s+\ime)^4}=\frac{\im\pi^2}{6(s-q^2)^2}
\end{equation}
which can be generalized to arbitrary integer powers $\geq3$
\begin{equation}\label{FI11}
   \int\dif^4 k\,\frac{1}{(k^2+2kq+s+\ime)^n}
   =\im\pi^2\frac{(n-3)!}{(n-1)!}\frac{1}{{(s-q^2)^{n-2}}}
\end{equation}
This can also be extended to \it{\ul{non-integral powers}}
\begin{equation}\label{FI12}
   \int\dif^4 k\,\frac{1}{(k^2+2kq+s+\ime)^n}
   =\im\pi^2\frac{\Gamma(n-2)}{\Gamma(n)}\frac{1}{{(s-q^2)^{n-2}}}
\end{equation}
and similarly
\begin{equation}\label{FI13}
   \int\dif^4 k\,\frac{k^\mu}{(k^2+2kq+s+\ime)^n}
   =-\im\pi^2\frac{\Gamma(n-2)}{\Gamma(n)}\frac{q^\mu}{{(s-q^2)^{n-2}}}
\end{equation}
\begin{equation}\label{FI14}
   \int\dif^4 k\,\frac{k^\mu k^\nu}{(k^2+2kq+s+\ime)^n}
   =\im\pi^2\frac{\Gamma(n-3)}{2\Gamma(n)}
   \Bigg[\frac{(2n-3)\,q^\mu q.^\nu}{{(s-q^2)^{n-2}}}
   +\frac{g^{\mu\nu}}{{(s-q^2)^{n-3}}}\Bigg]
\end{equation}

\bibliographystyle{C:/Sty/prsty}
%\bibliographystyle{spmpsci}
%\bibliography{RefIngvar,RefQED,RefDFT}

\begin{thebibliography}{1}

\bibitem{ILBook11}
I. Lindgren, {\em \textit{Relativistic Many-Body Theory: A New
  Field-Theoretical Approach}} (Springer-Verlag, New York, 2011).

\bibitem{LM86}
I. Lindgren and J. Morrison, {\em \textit{Atomic Many-Body Theory}} (Second
  edition, Springer-Verlag, Berlin, 1986, reprinted 2009).

\bibitem{DanJoh11}
D. Hedendahl and J. Holmberg,   (to be submitted to Phys. Rev. Lett.)  .

\bibitem{Adkins83}
G. Adkins, Phys. Rev. D {\bf 27},  1814  (1983).

\bibitem{Adkins86}
G. Adkins, Phys. Rev. D {\bf 34},  2489  (1986).

\bibitem{JohDan11}
J. Holmberg and D. Hedendahl,   (to appear in ArXiv:quant.ph)  .

\bibitem{PS95}
M.~E. Peskin and D.~V. Schroeder, {\em \textit{An introduction to Quantun Field
  Theory}} (Addison-Wesley Publ. Co., Reading, Mass., 1995).

\bibitem{MS84}
F. Mandl and G. Shaw, {\em \textit{Quantum Field Theory}} (John Wiley and Sons,
  New York, 1986).

\bibitem{Sak67}
J.~J. Sakurai, {\em \textit{Advanced Quantum Mechanics}} (Addison-Wesley Publ.
  Co., Reading, Mass., 1967).

\end{thebibliography}
%\end{thebibliography}

\end{document}